\documentclass{aa}
\usepackage[dvips]{graphicx}
\usepackage{txfonts}
\usepackage{aalongtable}
\begin{document}
\title{The Mira variable S~Ori: Relationships between 
the photosphere, molecular layer, dust shell, and SiO maser shell
at 4 epochs
\thanks{Based on observations made with the Very Large 
Telescope Interferometer (VLTI) at the 
Paranal Observatory under program IDs 074.D-0075
and 076.D-0338.}\fnmsep
\thanks{Based on observations made with the 
Very Long Baseline Array (VLBA) under project BB192. 
The VLBA is operated by the National Radio 
Astronomy Observatory (NRAO). The National
Radio Astronomy Observatory is a facility of the National
Science Foundation operated under cooperative
agreement by Associate Universities, Inc.}
\thanks{Tables \ref{tab:epochA_midi}--\ref{tab:epochC_comps} are
only available in electronic form at the CDS via anonymous ftp
to cdsarc.u-strasbg.fr (130.79.125.5) or via 
http://cdsweb.u-strasbg.fr/abstract.html}
\thanks{Color versions of Figs. 2--10,11,12 are available in
electronic form via http://www.edpsciences.org}
}
\author{
M.~Wittkowski\inst{1} \and
D.~A.~Boboltz\inst{2} \and
K.~Ohnaka\inst{3} \and
T.~Driebe\inst{3} \and
M.~Scholz\inst{4,5}
}
\institute{
European Southern Observatory, Karl-Schwarzschild-Str. 2,
85748 Garching bei M\"unchen, Germany, 
\email{mwittkow@eso.org}
\and
United States Naval Observatory, 3450 Massachusetts Avenue, 
NW, Washington, DC 20392-5420, USA 
\email{dboboltz@usno.navy.mil}
\and
Max-Planck-Institut f\"ur Radioastronomie, 
Auf dem H\"ugel 69, 53121 Bonn, Germany
\and
Institut f\"ur Theoretische Astrophysik der Univ. Heidelberg,
Albert-Ueberle-Str. 2, 69120 Heidelberg, Germany
\and
Institute of Astronomy, School of Physics,
University of Sydney, Sydney NSW 2006, Australia
}
\titlerunning{The Mira variable S~Ori: Photosphere, molecular layer, 
dust shell, and SiO maser shell}
\date{Received \dots; accepted \dots}
\abstract{}
{We present the first multi-epoch study that includes 
concurrent mid-infrared and radio interferometry of an
oxygen-rich Mira star.}
{We obtained mid-infrared interferometry of
S~Ori with VLTI/MIDI at four epochs in December 2004, 
February/March 2005, November 2005, and December 2005.
We concurrently observed $v=1, J=1-0$ (43.1\,GHz), 
and $v=2, J=1-0$ (42.8\,GHz) SiO maser emission toward S~Ori 
with the VLBA in January, February, and November 2005.
The MIDI data are analyzed using self-excited dynamic 
model atmospheres including molecular layers,
complemented by a radiative transfer model of the circumstellar dust shell. 
The VLBA data are reduced to the spatial structure and kinematics
of the maser spots.}
{The modeling of our MIDI data results in phase-dependent 
continuum photospheric angular diameters 
of 9.0\,$\pm$\,0.3\,mas (phase 0.42), 7.9\,$\pm$\,0.1\,mas (0.55), 
9.7\,$\pm$\,0.1\,mas (1.16), and 9.5\,$\pm$\,0.4\,mas (1.27).
The dust shell can best be modeled with Al$_2$O$_3$ grains using 
phase-dependent inner boundary radii between 1.8 and 2.4 photospheric radii. 
The dust shell appears to be more compact with greater optical depth near 
visual minimum ($\tau_V\sim$\,2.5), and more extended with lower optical 
depth after visual maximum ($\tau_V\sim$\,1.5).
The ratios of the 43.1\,GHz/42.8\,GHz SiO maser ring radii 
to the photospheric radii are
2.2\,$\pm$0.3/2.1\,$\pm$0.2 (phase 0.44), 
2.4\,$\pm$0.3/2.3\,$\pm$0.4 (0.55), 
and 2.1\,$\pm$\,0.3/1.9\,$\pm$\,0.2 (1.15).
The maser spots mark the region of the molecular 
atmospheric layers just beyond the steepest decrease in the 
mid-infrared model intensity profile. Their velocity structure 
indicates a radial gas expansion.}
{S~Ori shows significant phase-dependences of photospheric radii 
and dust shell parameters. Al$_2$O$_3$ dust grains and 
SiO maser spots form at relatively small radii of $\sim1.8-2.4$ 
photospheric radii.
Our results suggest increased mass loss and dust formation
close to the surface near the minimum visual phase, when Al$_2$O$_3$ dust 
grains are co-located with the molecular gas and the 
SiO maser shells,
and a more expanded dust shell after visual maximum.
Silicon does not appear to be bound in dust, as 
our data show no sign of silicate grains.}
\maketitle
\section{Introduction}
The evolution of cool luminous stars, including Mira variables, 
is accompanied by significant mass loss to the
circumstellar environment (CSE) with mass-loss rates of up
to $10^{-4}\ \mathrm{M}_\odot/\mathrm{year}$ 
(e.g., Jura \& Kleinmann \cite{jura90}).
This mass-loss process significantly affects any further stellar
evolution and is one of the most important sources for 
chemical enrichment of the interstellar medium.
The detailed nature of the mass-loss process from evolved stars, 
and especially its connection with the pulsation mechanism in 
the case of Mira variable stars, is a matter of current 
investigation.

The conditions close to the stellar surface can be studied well
by means of optical long-baseline interferometry.
This technique has provided information regarding the stellar 
photospheric diameter, effective temperature, 
center-to-limb intensity variation (CLV), and atmospheric molecular 
layers for a number of Mira variables
(see, e.g., 
Quirrenbach et al. \cite{quirrenbach92};
Haniff et al. \cite{haniff95}; 
van Belle et al. \cite{vanbelle96};
Thompson et al. \cite{thompson02}; 
Mennesson et al. \cite{mennesson02};
Hofmann et al. \cite{hofmann02}; 
van Belle et al. \cite{vanbelle02};
Perrin et al. \cite{perrin04};
Millan-Gabet et al. \cite{millan05}). 

Woodruff et al. (\cite{woodruff04}) and Fedele et al. 
(\cite{fedele05}) have recently shown that observed near-infrared
$K$-band visibilities of the oxygen-rich prototype Mira 
variables \object{$o$~Cet} and \object{R~Leo} are very 
different from uniform disc (UD) models already in the first lobe 
of the visibility function, and that they correspond closely to predictions 
by self-excited dynamic Mira model atmospheres that include 
effects from molecular layers  
(models by Hofmann et al. \cite{hofmann98}; Tej et al. \cite{tej03b}; 
Ireland et al. \cite{ireland04a,ireland04b}).
Ohnaka et al. (\cite{ohnaka06b}) studied the comparison of 
the same dynamic Mira model atmospheres with {\it mid-infrared} 
interferometric and spectroscopic observations.
Recently, Ireland \& Scholz (\cite{ireland06}) added  
the formation of dust in a 
self-consistent way
to the same dynamic 
Mira model atmospheres. They find that dust would form at 2--3 times 
the average photospheric radius for certain plausible 
parameter values.

The structure of the atmospheric molecular shells located above 
the photosphere, as well as the dust shell, can be probed by 
mid-infrared interferometry. 
This has been successfully demonstrated by Ohnaka et al. 
(\cite{ohnaka05}) using the spectro-interferometric 
capabilities of the VLTI/MIDI facility for observations of 
the oxygen-rich Mira star \object{RR~Sco}. 
The model used in this work includes a warm molecular
layer consisting of SiO and H$_2$O, as well as an 
optically-thin dust shell of Al$_2$O$_3$ and silicate.
Recent VLTI/MIDI observations of the carbon-rich 
Mira star \object{V~Oph} by Ohnaka et al. (\cite{ohnaka07})
indicate that carbon-rich Miras
also have extended atmospheric layers of polyatomic 
molecules (C$_2$H$_2$) and dust shells (amorphous carbon and SiC).
Information on dust shells around Mira variables has 
also been derived using mid-infrared interferometry with the
Berkeley Infrared Spatial Interferometer ISI 
(see e.g., Danchi et al. \cite{danchi94}; 
Weiner et al. \cite{weiner06}; Tatebe et al. \cite{tatebe06}). 
Ireland et al. (\cite{ireland05}) recently used 0.9\,$\mu$m 
interferometric polarimetry measurements of the Mira 
variables \object{R~Car} and RR~Sco to place 
constraints on the distribution of the dust shell. 

Complementary information regarding the molecular shells around
oxygen-rich AGB stars can be obtained by observing 
the maser radiation that some of these molecules emit. 
Maser emission from the three most common maser molecules, 
SiO, H$_2$O, and OH, traces regions of the CSE on scales 
from a few to several hundred AU. Masers 
provide a unique probe of structure, kinematics, and 
polarization properties of the environment of these stars. 
The SiO maser radiation in the CSE of oxygen-rich Mira variables
has been mapped using the Very Long Baseline Array (VLBA) by, 
e.g., Diamond et al. (\cite{diamond94}),
Kemball \& Diamond (\cite{kemball97}),
Boboltz et al. (\cite{boboltz97}),
Diamond \& Kemball (\cite{diamond03}),
Cotton et al. (\cite{cotton04}), 
Boboltz \& Wittkowski (\cite{boboltz05}, hereafter BW05), 
Boboltz \& Diamond (\cite{boboltz05b}), and Cotton et al. (\cite{cotton06}).
The SiO maser emission is typically found to arise from a clumpy ring 
within a few stellar radii of the photosphere, 
indicating a tangential amplification process. 
Humphreys et al. (\cite{humphreys96,humphreys02}) 
have predicted the location at which SiO molecules form in the CSE of 
Mira variables by combining maser models with hydrodynamic 
pulsation models. Their model predictions are roughly 
consistent with the measurements mentioned above.

Observational results regarding the detailed relationships 
between the stellar photosphere, the molecular layer, the dust 
shell, and the SiO maser ring often suffer from uncertainties 
inherent in comparing observations of variable stars widely 
separated in time and stellar phase, as discussed in BW05.
To overcome this limitation, we have established
a program of concurrent infrared interferometry using 
the VLTI and radio interferometry at the VLBA. The former 
aims at constraining the photospheric radius, the 
characteristics of atmospheric molecular layers, and parameters of 
the dust shell. The latter aims at concurrently mapping 
the SiO maser emission. Our final goal is 
a better understanding of the mass-loss process and its connection 
to stellar pulsation.
Our pilot study in BW05 on the 
Mira variable S~Ori included coordinated near-infrared $K$-band 
interferometry to constrain the stellar photospheric diameter 
and VLBA mapping of the SiO maser radiation toward this source.

A further uncertainty in comparing photospheric radii to the 
extensions of the dust and maser shells arises from the 
complication that near- and mid-infrared CLVs of finite 
bandwidth include a blend of intensities from 
continuum-forming layers and overlying molecular layers. 
This effect has often resulted in over-estimated continuum photospheric 
diameter values (see, e.g. the discussions in 
Jacob \& Scholz \cite{jacob02}; 
Mennesson et al. \cite{mennesson02}; 
van Belle et al. \cite{vanbelle02}; Woodruff et al. \cite{woodruff04}; 
Ireland et al. \cite{ireland04a,ireland04b}; 
Perrin et al. \cite{perrin04}; Fedele et al. \cite{fedele05}). 
An overestimated photospheric diameter would result in 
biased relative distances of the dust shell and the 
maser ring from the stellar surface, even if obtained at 
the same stellar phase and cycle. A detailed comparison of 
observations to dynamic model atmospheres, as mentioned above, 
can be used to relate the observable quantities to the continuum
photospheric radius and thus to overcome this limitation.

Here, we present VLTI/MIDI mid-infrared interferometry of S~Ori
at 4 epochs/stellar phases and coordinated VLBA mapping of 
the SiO 42.8\,GHz and 43.1\,GHz maser transitions at 3 
epochs/stellar phases that are contemporaneous 
to the first 3 of our 4 MIDI epochs.
The first two epochs are located near the stellar minimum, 
and the later epochs shortly after the following stellar maximum.
\section{Lightcurve and characteristics of S~Ori}
\label{sec:lightcurve}
\begin{figure}
\centering
\resizebox{1\hsize}{!}{\includegraphics{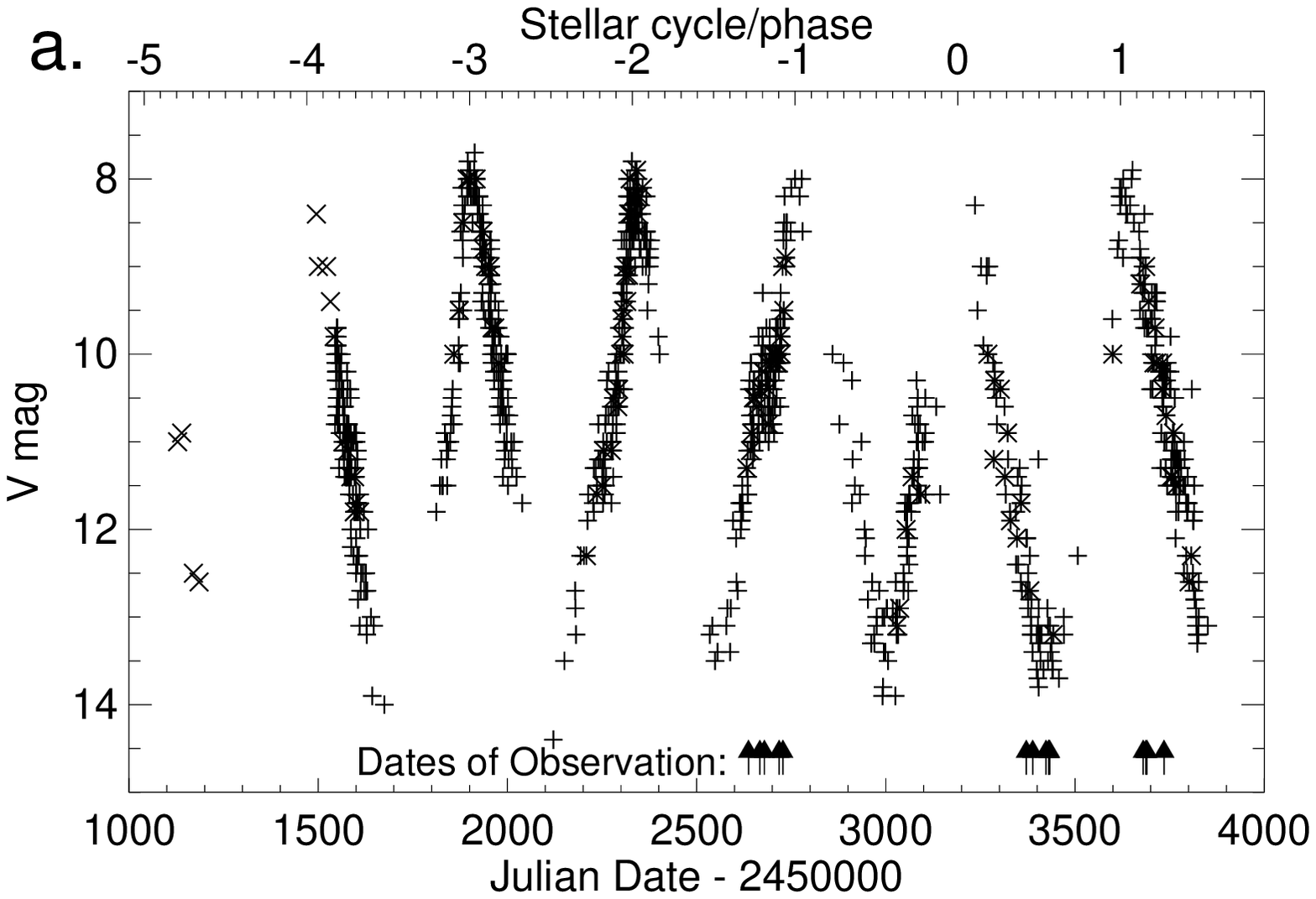}}

\resizebox{1\hsize}{!}{\includegraphics{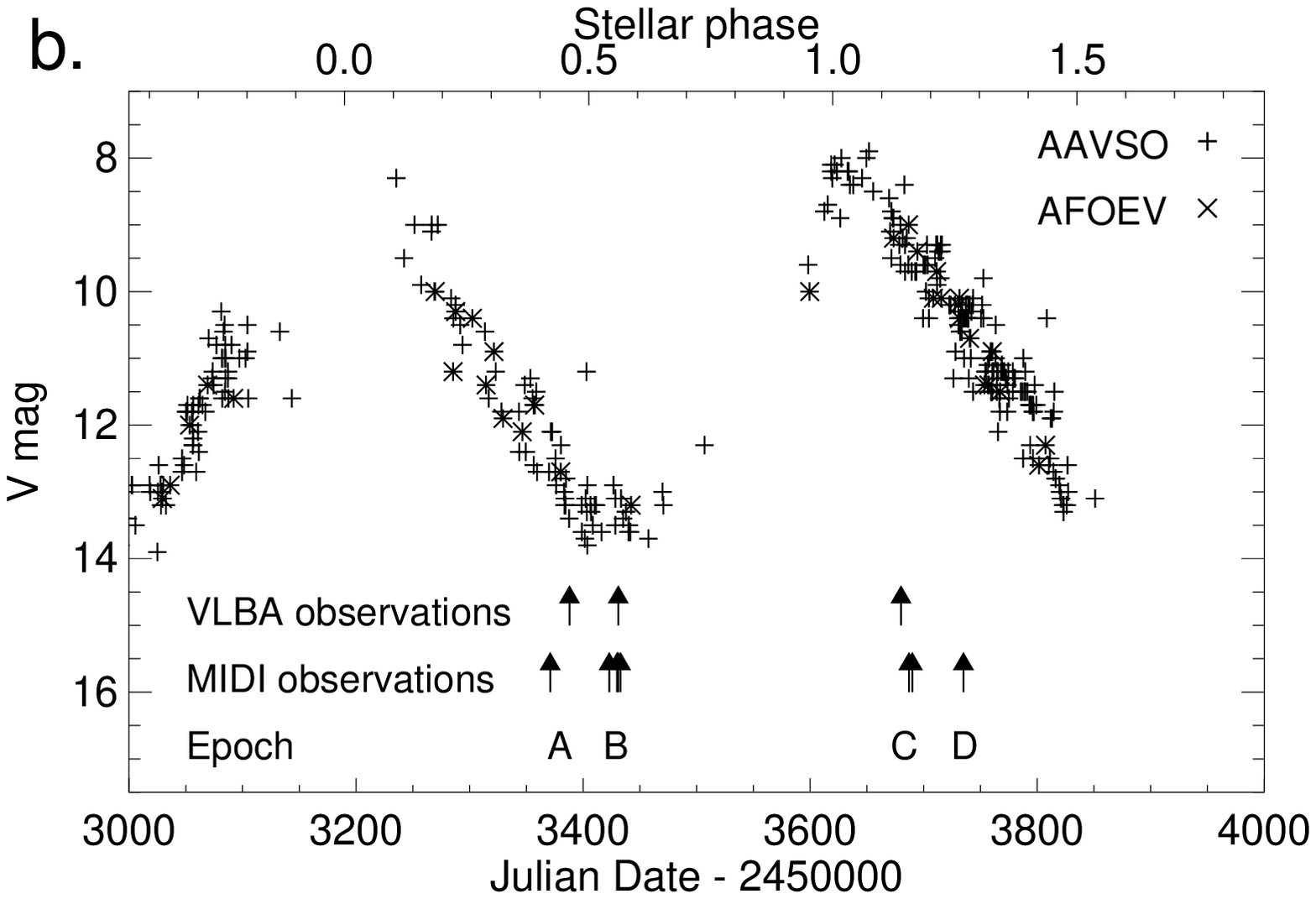}}
\caption{Visual lightcurve of S~Ori as a function of Julian Date
and stellar cycle/phase. Data are from the AAVSO (Henden et
al. \cite{henden06}) and AFOEV (CDS) databases. We adopt a 
period of $P=430\ \mathrm{days}$ and JD of last maximum 
brightness $T_0=2453190$.
The arrows indicate the dates of our VLTI and VLBA observations.
a) The most recent 5--6 cycles illustrating the correspondence of
the data with our adopted $P$ and $T_0$ values. The indicated 
dates of observations are the epochs from BW05 (JD 
between 2600 and 2800) and the epochs of the present paper 
(JD between 3300 and 3800).
b) Enlarged view covering our VLTI/MIDI and 
VLBA observations presented in this paper. The dates of 
observations are combined into 4 epochs A, B, C, D.
}
\label{fig:lightcurve}
\end{figure}
S~Ori is a Mira variable star with spectral type M6.5e--M9.5e
and $V$ magnitude 7.2--14.0 (Samus et al. \cite{samus04}).
Templeton et al. (\cite{templeton05}) 
report that S~Ori's period appears to vary in 
a seemingly sinusoidal fashion between about 400 and 450 days
over about the past 100 years. 
The mean period is $\overline{P}=420.0\ \mathrm{days}$ with
a latest period, corresponding to our observation epochs, of 
$P\approx\ 430\ \mathrm{days}$. We find a good correspondence 
with the AAVSO (Henden et al. \cite{henden06}) and AFOEV (CDS) 
data for the cycle of our observations, as well as for the most 
recent $\sim$\,5 cycles using $P=430\ \mathrm{days}$ and a 
Julian Date of last maximum brightness 
$T_0=2453190\ \mathrm{days}$. Figure~\ref{fig:lightcurve} shows 
the visual lightcurve of S~Ori as a function of Julian Date 
and stellar phase based on these values. Also indicated are 
the dates/epochs of our VLTI and VLBA observations as described 
in Sects.~\ref{sec:midiobs} and \ref{sec:vlbaobs}.

The distance to S~Ori, as for many Mira stars, is not well 
known. Van Belle et al. (\cite{vanbelle02}) have estimated the 
distance to S~Ori to
$480\ \mathrm{pc}\ \pm\ 120\ \mathrm{pc}$ based on a 
calibration of the period-luminosity relationship by
Feast et al. (\cite{feast89}), and we adopt this value. 

The near-infrared $K$-band UD 
angular diameter $\Theta^K_\mathrm{UD}$ of S~Ori has been 
measured by van Belle et al. (\cite{vanbelle96}),
Millan-Gabet et al. (\cite{millan05}), and by BW05
to values between 9.6\,mas and 10.5\,mas at different
phases.
Diameter measurements of S~Ori at other wavelengths have so 
far not been reported.

S~Ori exhibits SiO and OH maser emission 
(Benson et al. \cite{benson90}), while a detection 
of H$_2$O maser 
toward S~Ori has not been reported. The maps of the 
SiO maser emission in BW05 are the first toward S~Ori.
They derived average distances and standard deviations
of the SiO maser spots from the center
of their distribution at phase $0.73$ of 
$9.4 \pm 1.4\ \mathrm{mas}$ and
$8.8 \pm 1.7\ \mathrm{mas}$ for the $43.1\ \mathrm{GHz}$ 
and $42.8\ \mathrm{GHz}$ transitions, respectively. 

Sloan \& Price (\cite{sloan98}) report on a relatively low 
dust-emission coefficient (DEC, the total emission of the dust
to the total emission of the star in the wavelength range 
7.7--14.0\,$\mu\mathrm{m}$) for S~Ori of $0.24$, i.e. similar 
to RR~Sco (DEC=0.21) or R~Leo (DEC=0.23), indicating
an optically thin dust shell.
\section{VLTI/MIDI measurements}
\label{sec:vlti}
\begin{table*}
\caption{VLTI/MIDI observations of S~Ori.}
\label{tab:vltiobs}
\begin{tabular}{lllrrlrllrrrr}
\hline\hline
Ep. & Date & Time  & JD & $\Phi_\mathrm{vis}$ & Conf. & $B$ & Disp. & BC & $B_p$ & PA & Seeing & $\tau_0$ \\
    &      & UTC  & -2.4E6 &                     &       &  &Elem. &     & [m]   & [deg]& [\arcsec] & [msec] \\\hline
A   & 2004-12-31 & 02:43 & 53370.6 & 0.42 & UT3-UT4 & 62\,m & Prism & {\tt HS} & 62.27 & -71.9 & 0.76 & 3.4 \\[1ex]
B   & 2005-02-21 & 02:37 & 53422.6 & 0.54 & UT2-UT4 & 89\,m & Prism & {\tt HS} & 74.31 &  83.0 & 0.74 & 3.6 \\
B   & 2005-03-01 & 02:39 & 53430.6 & 0.56 & UT3-UT4 & 62\,m & Prism & {\tt HS} & 41.65 & -55.6 & 0.62 & 3.6 \\
B   & 2005-03-03 & 00:59 & 53432.5 & 0.56 & UT2-UT4 & 89\,m & Prism & {\tt HS} & 84.35 &  82.7 & 0.98 & 3.7 \\[1ex]
C   & 2005-11-12 & 08:57 & 53686.9 & 1.16 & UT1-UT4 &130\,m & Prism & {\tt SP} &123.02 &  62.8 & 0.54 & 4.4 \\
C   & 2005-11-15 & 07:21 & 53689.8 & 1.16 & UT2-UT3 & 47\,m & Prism & {\tt SP} & 46.26 &  43.9 & 0.61 & 2.9 \\[1ex]
D   & 2005-12-30 & 03:00 & 53734.6 & 1.27 & UT1-UT4 &130\,m & Prism & {\tt SP} &123.89 &  59.4 & 1.12 & 2.8 \\\hline
\end{tabular}
\end{table*}
%
\label{sec:midiobs}
We obtained mid-infrared (8--13\,$\mu$m) interferometry 
of S~Ori with the instrument MIDI (Leinert et al.
\cite{leinert03}) at the ESO VLT Interferometer (VLTI; 
Glindemann et al. \cite{glindemann03}) in service mode between 
31 December 2004 and 30 December 2005. Light was combined from 
two 8\,m Unit Telescopes (UTs). We chose to use the prism with 
spectral resolution $R=\lambda/\Delta\lambda=30$ to disperse 
the interferograms. Two various techniques of beam 
combinations were used, namely {\tt HIGH\_SENS} ({\tt HS}, 2004-12-31 
to 2005-03-03) and {\tt SCI\_PHOT} ({\tt SP}, 2005-11-12 to 2005-12-30) 
combinations. Using the {\tt HS} combination, all arriving light is 
first combined interferometrically, and the photometric spectrum 
for each incoming beam is recorded sequentially. Using the
{\tt SP} combination, beam splitters are used to record the 
interferograms and the photometric spectra simultaneously. 
An estimate of the photometric count rates at the time of 
recording the interferometric data is needed to compute the 
raw visibility values, and the {\tt SP} technique is thus expected to 
result in a higher precision of the visibility measurement for bright targets. 
Our observations followed the standard procedures for 
MIDI service mode observations as described in detail in the 
MIDI user manual\footnote{http://www.eso.org/instruments/midi}. 
For principles of observations with MIDI see 
also Leinert et al. (\cite{leinert04}).
  
The details of our observations are listed in 
Table~\ref{tab:vltiobs}. Listed are the 
epoch, date, time, and Julian Date (JD) of the observation; 
the visual phase $\Phi_\mathrm{vis}$; 
the baseline configuration used with its ground length $B$; 
the dispersive element and beam combiner BC 
({\tt HS} for {\tt High\_Sens} and {\tt SP} for {\tt Sci\_Phot}); 
and the projected 
baseline length $B_p$ and its position angle PA on sky (deg. 
east of north). The last two columns describe the ambient 
conditions DIMM seeing, and coherence time $\tau_0$, both 
at 500\,nm.
Dates of observation were combined into 4 epochs A, B, C, D.
Each epoch has a width of less than 17 days, i.e. less than  
5\% of S~Ori's variability period. The dates of observation and
the epochs compared to the visual 
lightcurve are indicated in Fig.~\ref{fig:lightcurve}.
We estimate the uncertainty in assigning a visual phase value 
to each of our observing epochs to $\sim 0.1$, taking into 
account both the time widths of our epochs and the uncertainty in 
the choice of S~Ori's current variability period and 
date of last maximum (see Sect.~\ref{sec:lightcurve}). 
Epochs A and B are located around stellar minimum with a 
separation of about 2 months (phase difference $\sim$\,0.1), 
while epochs C and D occurred about a year later
shortly after the following stellar maximum with a separation
of 6 weeks.
Additional MIDI/UT data on S~Ori from 2004-12-29, 2005-01-02,
2005-02-28, 2006-03-12, 2006-03-12, as well as MIDI/AT data
taken between October 2005 and March 2006 were not used because
of various technical problems with these data sets.
The interferometric field of view of MIDI used with the UTs
is $\sim$\,250\,mas, which corresponds to about 25 times
the expected photospheric angular diameter of S~Ori.

We used \object{$\theta$~CMa} (HD\,50778) as the interferometric 
calibrator, a K4 giant at a distance on sky of 27$^\circ$, which
was observed within about 30 minutes of each S~Ori observation.
A number of observations of other MIDI calibration stars using 
the same instrument settings were obtained during some of our
observation nights for other programs, and we make use of these
data as well. Table~\ref{tab:midical} lists the calibration stars
and their properties. These parameters are taken 
from the instrument consortium's 
catalog\footnote{http://www.ster.kuleuven.ac.be/$\sim$tijl/MIDI\_calibration/mcc.txt}. 
Listed are the HD number, the name, the
spectral type, the limb-darkened diameter $\Theta_\mathrm{LD}$, 
the effective temperature $T_\mathrm{eff}$, the 
flux at 12\,$\mu$m, and the distance on sky from S~Ori $\zeta$.
The UD diameter for K giants in the mid-infrared
based on spherical model atmospheres is smaller by $\sim$\,5\% 
than $\Theta_\mathrm{LD}$ (estimated with procedures as used in 
Wittkowski et al. \cite{wittkowski06}).
\object{HD~50778} is our main calibrator. The other calibrators 
were observed during some of our observation nights for other 
programs. Among the interferometric calibration stars listed 
in Table~\ref{tab:midical},
absolutely calibrated spectra are available for HD~31421 and 
HD~61935 (Cohen et al. \cite{cohen99}).
\subsection{MIDI data reduction}
\label{sec:midireduction}
\begin{table}
\caption{Properties of the MIDI calibration stars used.}
\label{tab:midical}
\begin{tabular}{rrrccrr}
\hline\hline
HD&Name& Sp.T.&$\Theta_\mathrm{LD}$ & $T_\mathrm{eff}$ & $F_{12}$ & $\zeta$\\ 
       &                 &           & [mas]    & [K]& [Jy] & [\degr]\\\hline
 50778 & $\theta$~CMa &   K4\,III & 3.90$\pm$0.02 & 4049 & 24.6 & 27\\[1ex]
 31421 & \object{$o^2$~Ori}    &   K2\,III & 2.65$\pm$0.01 & 4500 & 13.6 & 12\\
 49161 & \object{17~Mon}       &   K4\,III & 2.44$\pm$0.01 & 4248 &  7.2 & 20\\
 61935 & \object{$\alpha$~Mon} &   G9\,III & 2.24$\pm$0.01 & 4813 & 10.2 & 36\\
 94510 & \object{u~Car}        &   K1\,III & 2.23$\pm$0.01 & 4951 &  6.9 & 90\\
107446 & \object{$\epsilon$~Cru}& K3.5\,III & 4.43$\pm$0.02 & 4139 & 32.4 &101 \\
\hline
\end{tabular}
\end{table}
Mean coherence factors and photometric count rates were obtained
for each data set of S~Ori and the calibration stars
using the {\tt MIA+EWS} software package, version 1.4 
(Jaffe, Koehler, Cotton, Chesneau, et al.\footnote{http://www.strw.leidenuniv.nl/$\sim$koehler/MIDI}),
complemented by the tool {\tt MyMidiGui}, 
version 1.41 (Hummel\footnote{http://www.sc.eso.org/$\sim$chummel/midi/midi.html}).
Data for 32 spectral channels within the nominal
MIDI wavelength range from 8.0\,$\mu$m to 13.0\,$\mu$m were 
used. The detector masks were calculated by the procedure 
of {\tt MyMidiGui}, and were used for both the {\tt MIA} and 
the {\tt EWS} analysis. The results obtained by {\tt MIA} and {\tt EWS} 
were found to be virtually identical for all datasets. We chose to
use in the following the results based on the {\tt EWS} analysis.
It is based on a coherent integration of the fringe signal 
after numerical compensation of optical path differences in 
each scan.

Calibrated visibility spectra for S~Ori were derived
using the transfer function obtained for our calibrator HD~50778. 
The scatter of all obtained transfer functions during the night was used to 
estimate its uncertainty. The latter depends on the sample 
of calibrations stars available per night, and we assumed a 
minimum uncertainty of the transfer function of 0.1 based on 
nights when many values were obtained. The total error of our 
calibrated S~Ori visibility spectrum includes the errors of 
the raw coherence factors computed by {\tt EWS}, the uncertainty of 
the calibrator diameters, and the scatter of the transfer 
function during the night. The total error is dominated by 
systematic uncertainties for the height of the interferometric 
transfer function, so that the relative accuracy of the
visibility values among the spectral channels is better 
than indicated by our total error bars.

We absolutely calibrated the photometric spectrum of S~Ori for
each data set using the interferometric calibrator HD~50778 also 
as the photometric calibration star.
The absolutely calibrated photometric spectrum of HD~50778 was
not available directly. We used instead the average of the 
available spectra of \object{HD~20644} and \object{HD~87837} 
(Cohen et al. \cite{cohen99}), scaled with the 
IRAS 12\,$\mu$m flux (Beichmann et al. \cite{beichmann88})
to the level of HD~50778. These stars 
have the same spectral type (K4\,III) and very similar effective 
temperatures as HD~50778 (4058\,K and 4094\,K versus 4049\,K).
In addition, we calibrated the 
photometric spectrum of HD~50778 using the spectro-photometric 
calibrator HD~31421, which was observed close in time during 
the night 2005-03-03, and verified that our 
synthetic HD~50778 spectrum is valid. 
We obtained integrated 8.0--13.0\,$\mu$m S~Ori flux densities for our 
MIDI epochs A to D of $187\ \pm\ 12\ \mathrm{Jy}$, 
$152\ \pm\ 17\ \mathrm{Jy}$, $198\ \pm\ 18\ \mathrm{Jy}$, and
$203\ \pm\ 13\ \mathrm{Jy}$. For comparison, the integrated 
IRAS LRS flux density for the same bandpass (8.0--13.0\,$\mu$m) 
is 169\,Jy.

The MIDI calibrated visibility, as well as calibrated flux values,
are detailed in Table~\ref{tab:epochA_midi}.
\subsection{MIDI data reduction results}
As an alternative and probably more intuitive quantity compared
to the visibility values, 
we calculated the equivalent UD diameters. The 
equivalent UD diameter was computed for each projected 
baseline length and spectral channel separately and defined 
as the diameter of that uniform disc that results in the same 
visibility value at each separate data point. Note that the 
true intensity profile is not expected to be a uniform disc. 
As a result, the equivalent UD diameters for 
different projected baseline lengths are not expected to 
coincide (not even for a circularly symmetric intensity
profile). However, these values are fully equivalent to the 
calibrated visibility values and can provide a first rough 
estimate of the characteristic size of the target. 

\begin{figure*}
\centering
\sidecaption
\vspace*{1cm}%
\includegraphics[width=8.8cm]{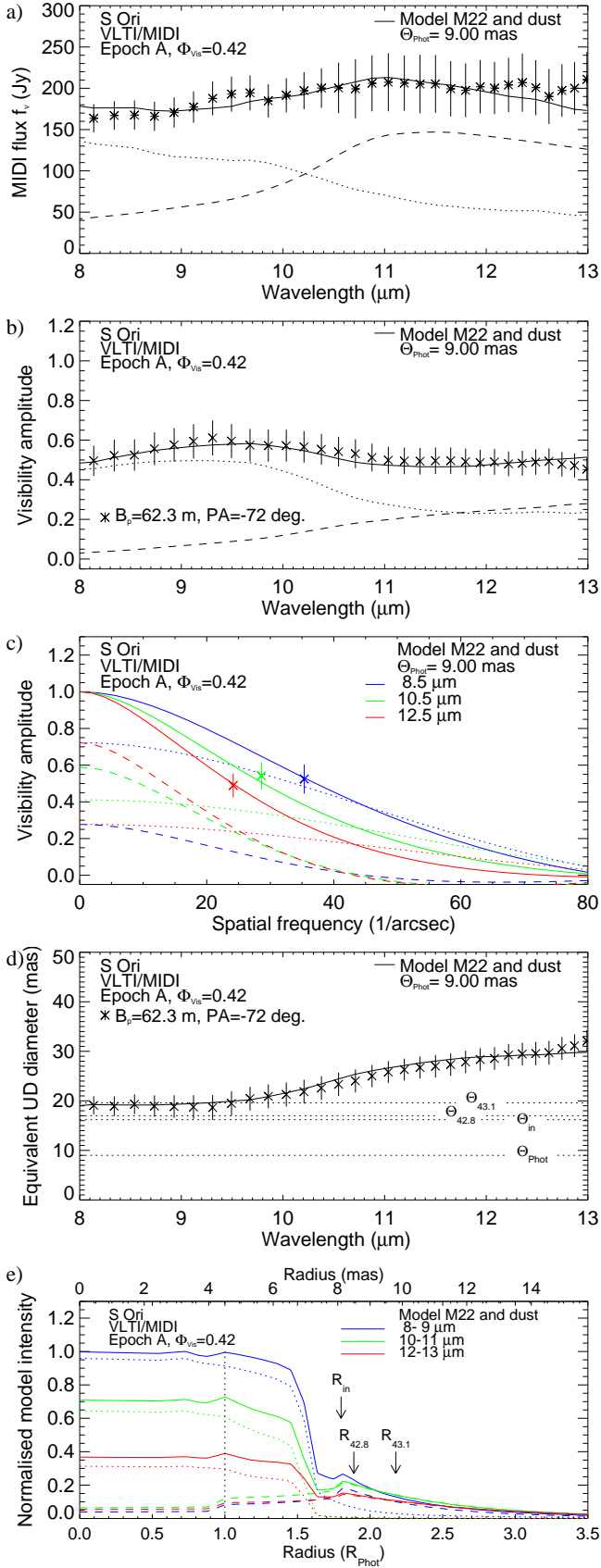}
\caption{
Results of our VLTI/MIDI 8--13\,$\mu$m interferometry
of S~Ori at epoch A, stellar phase 0.42. 
Shown are our MIDI data compared to the best-fitting model
with parameters listed in Table~\protect\ref{tab:midiresults}.
In each panel, the {\protect\tt\sf x}-symbols with error bars denote the 
measured values, the solid lines the synthetic values of the 
overall model consisting of the stellar model atmosphere and the
dust shell. The dotted lines denote the contribution from the
attenuated atmosphere model alone, and the dashed lines the contribution of 
only the dust shell (attenuated input radiation subtracted).\protect\\
a) The calibrated MIDI flux spectrum.\protect\\
b) The calibrated MIDI visibility values 
as a function of wavelength.\protect\\
c) The calibrated MIDI visibility values as a function of 
spatial frequency for the example of three spectral channels 
at wavelengths 8.5\,$\mu$m, 10.5\,$\mu$m, and 
12.5\,$\mu$m.\protect\\
d) The equivalent UD diameter values. Also indicated 
are the photospheric angular diameter $\Theta_\mathrm{Phot}$ and
the inner dust shell boundary $\Theta_\mathrm{in}$
as obtained by our model fit, as well as 
the mean 42.8\,GHz and 43.1\,GHz SiO ring angular 
diameters (Sect.~\protect\ref{sec:vlba}).\protect\\
e) The model intensity profiles at spectral bands 8--9\,$\mu$m, 
10--11\,$\mu$m, and 12--13\,$\mu$m. The bottom abscissa shows 
the radius as a function of the photospheric radius $R_\mathrm{Phot}$, 
and the top abscissa the corresponding angular scale obtained 
from our model fit. The arrows indicate 
the location of the inner boundary radius of the dust 
shell $R_\mathrm{in}$, as well as of the mean 
43.1\,GHz and 42.8\,GHz SiO maser spot radii $R_{43.1}$ and $R_{42.8}$.
}
\label{fig:midiA}
\end{figure*}

Figures~\ref{fig:midiA}--\ref{fig:midiD}
show, for MIDI epochs A--D, 
respectively, the obtained MIDI flux spectra, the MIDI 
visibility values as a function of wavelength, the MIDI 
visibility values as a function of spatial frequency for 
the example of three spectral channels and the equivalent
UD diameter values. Also shown are the synthetic 
visibility and flux values of the atmosphere and dust models 
described below in Sect.~\ref{sec:midimodel} in each panel, 
as well as the intensity profiles of these models for the 
example of three bandpasses in the bottom panels. 

The general shape of our S~Ori visibility functions, or 
equivalent UD diameter values, in 
Figs.~\ref{fig:midiA}--\ref{fig:midiD} is qualitatively similar 
to the MIDI data of the Mira variable RR~Sco described 
in Ohnaka et al. (\cite{ohnaka05}). Objects such as these are
characterized by a partially resolved stellar disc, including 
atmospheric molecular layers that are optically thick at mid-infrared 
wavelengths, surrounded by a spatially resolved, optically 
thin dust shell. The visibility increases between about 
8--9\,$\mu$m, corresponding to a quasi-constant equivalent 
UD diameter that is roughly twice the continuum 
photospheric size. Here, the observed intensity is 
dominated by radiation from optically thick molecular layers at 
distances up to roughly two continuum photospheric radii. 
Beyond about 9--10\,$\mu$m, the visibility function flattens, 
corresponding to an increasing equivalent UD diameter.
In this wavelength region, spatially resolved, optically thin 
radiation from the dust shell starts to represent a 
considerable part of the total intensity. In addition, 
extinction of the stellar light by the dust shell becomes 
important. At the longest wavelengths of MIDI, spatially
resolved radiation from the dust shell dominates the measured 
intensity.
A detailed model of S~Ori's atmosphere and dust shell is 
described in Sects.~\ref{sec:midimodel} 
and \ref{sec:midimodelresults}.
\subsection{MIDI data modeling}
\label{sec:midimodel}
Mid-infrared photometric and interferometric data of Mira 
stars are in general sensitive to details of the structure 
of the stellar atmosphere in which molecular layers of 
various geometrical and optical characteristics (e.g., Tej 
et al. \cite{tej03a}) lie above the continuum-forming 
photosphere, as well as to the surrounding dust shell 
(cf. e.g., Ohnaka et al. \cite{ohnaka05,ohnaka06b}; 
Ireland \& Scholz \cite{ireland06};
Weiner et al. \cite{weiner06}). We used the P and M model
atmosphere series (Ireland et al. \cite{ireland04a,ireland04b}) 
to describe the stellar atmosphere including the continuum 
photosphere and overlying molecular layers, and added an 
ad-hoc radiative transfer model of the dust shell. The details 
of our modeling are described in the following.  

\subsubsection{Atmosphere model for S~Ori}
Few dynamic atmosphere models for oxygen-rich Mira stars
are available that include the effects of molecular 
layers (see, e.g., Scholz \cite{scholz01,scholz03}). The 
P and M model series (Ireland et al. 
\cite{ireland04a,ireland04b}) 
are complete self-excited dynamic model atmosphere of Mira 
stars, and they have successfully explained a lot of observational 
data (e.g.; Hofmann et al. \cite{hofmann00,hofmann01,hofmann02}; 
Woodruff et al. \cite{woodruff04}; 
Fedele et al. \cite{fedele05}).
These model series have been constructed to match the prototype 
oxygen-rich Mira stars $o$~Cet and R~Leo. 
The P and M series differ with respect to the mass of the 
so-called ``parent star'', which is the hypothetical 
non-pulsating equivalent of the pulsating Mira variable.
The geometric pulsation of the Mira occurs around the parent 
star's Rosseland radius $R = R(\tau_\mathrm{Ross}=1)$ 
(Ireland et al. \cite{ireland04a,ireland04b}).
The parent star has solar metallicity, luminosity
$L/\mathrm{L}_\odot$=3470, period 332 days,
mass $M/\mathrm{M}_\odot$=1.0 (P series) and 1.2 (M),
radius $R_p/\mathrm{R}_\odot$=241 (P) and 260 (M)
(Hofmann et al. \cite{hofmann98}). The moderately
larger radius of the M-series parent star leads to a slightly 
lower effective temperature for the parent star and 
systematically lower phase-dependent effective temperatures 
for the pulsating Mira compared to the P series. 
The M models tend to exhibit less pronounced cycle-to-cycle 
variations than the P models and to have more compact 
atmospheres (Ireland et al. \cite{ireland04b}).
Effects on interferometric diameter measurements resulting from
the differences between the P and M model series are subtle.
When comparing P and M models of similar phases to near-infrared
interferometric data of R~Leo, the resulting 
Rosseland angular diameter values were found to agree 
within $\la$\,1.5\% (Fedele et al. \cite{fedele05}).

Compared to $o$~Cet and R~Leo, for which the
P and M model series are designed, S~Ori is a slightly 
cooler Mira variable (M6.5--M9.5 versus M5--M9/M6--M9.5, 
Samus et al. \cite{samus04}) with a longer
period (420\,d versus 332\,d/310\,d,
Samus et al. \cite{samus04}), a larger main sequence precursor 
mass (1.3\,M$_\odot$ versus 1.0\,M$_\odot$/1.2\,M$_\odot$,
Wyatt \& Cahn \cite{wyatt83}), and a larger radius 
(mean continuum photospheric radii roughly $R\sim$450\,R$_\odot$ versus 
$\sim$350\,R$_\odot$; based on 
BW05, Woodruff et al. \cite{woodruff04}, Fedele et al. \cite{fedele05}).
However, when scaled to variability phases between 0 and 1 and
to the corresponding angular size on sky, the general model 
results are not expected to be dramatically different for 
S~Ori compared to Mira stars, such as $o$~Cet and R~Leo. 
Hereby, the M series is expected to be better-suited 
to describing S~Ori than the P series because of its 
higher parent-star mass.
We concluded that the use of the M series is currently
the best available option for describing S~Ori's atmosphere 
including the continuum photosphere and overlying atmospheric 
molecular layers.
\begin{figure}
\centering
\resizebox{1.0\hsize}{!}{\includegraphics{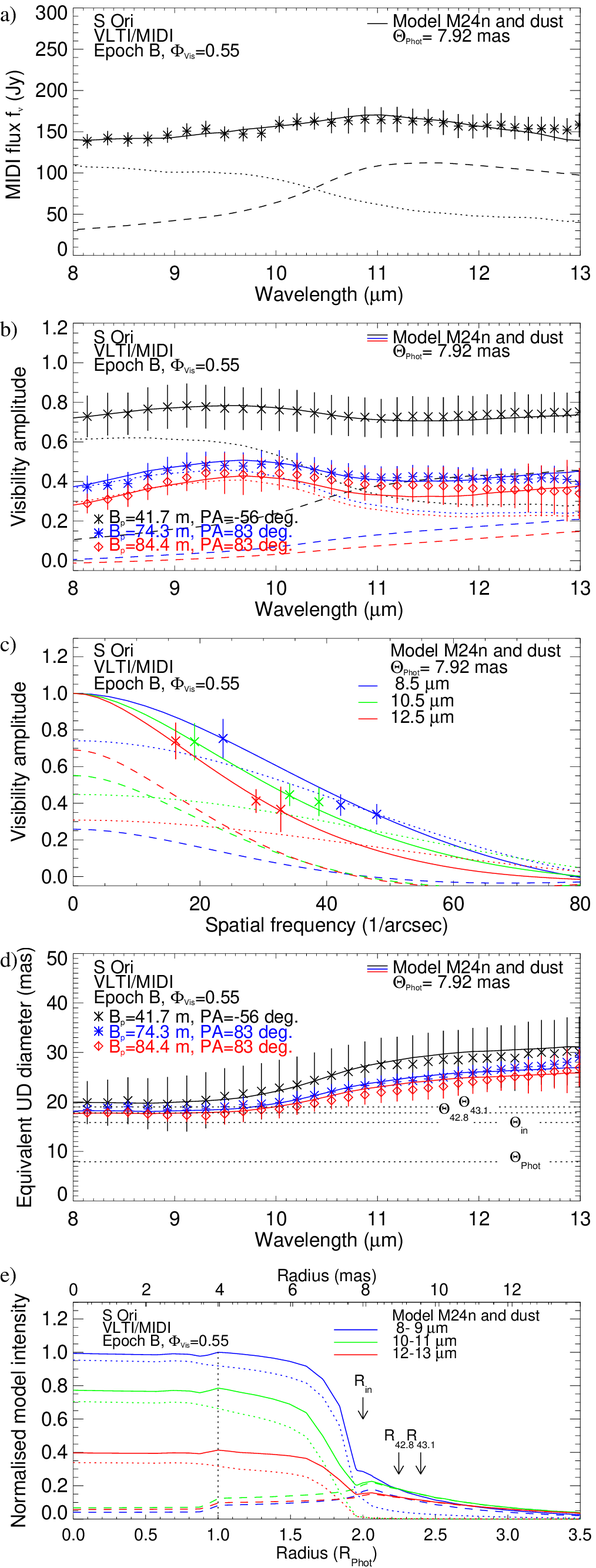}}
\caption{Results of our VLTI/MIDI 8--13\,$\mu$m interferometry
of S~Ori at epoch B, stellar phase 0.55. As 
Fig.~\protect\ref{fig:midiA}.
}
\label{fig:midiB}
\end{figure}
\begin{figure}
\centering
\resizebox{1.0\hsize}{!}{\includegraphics{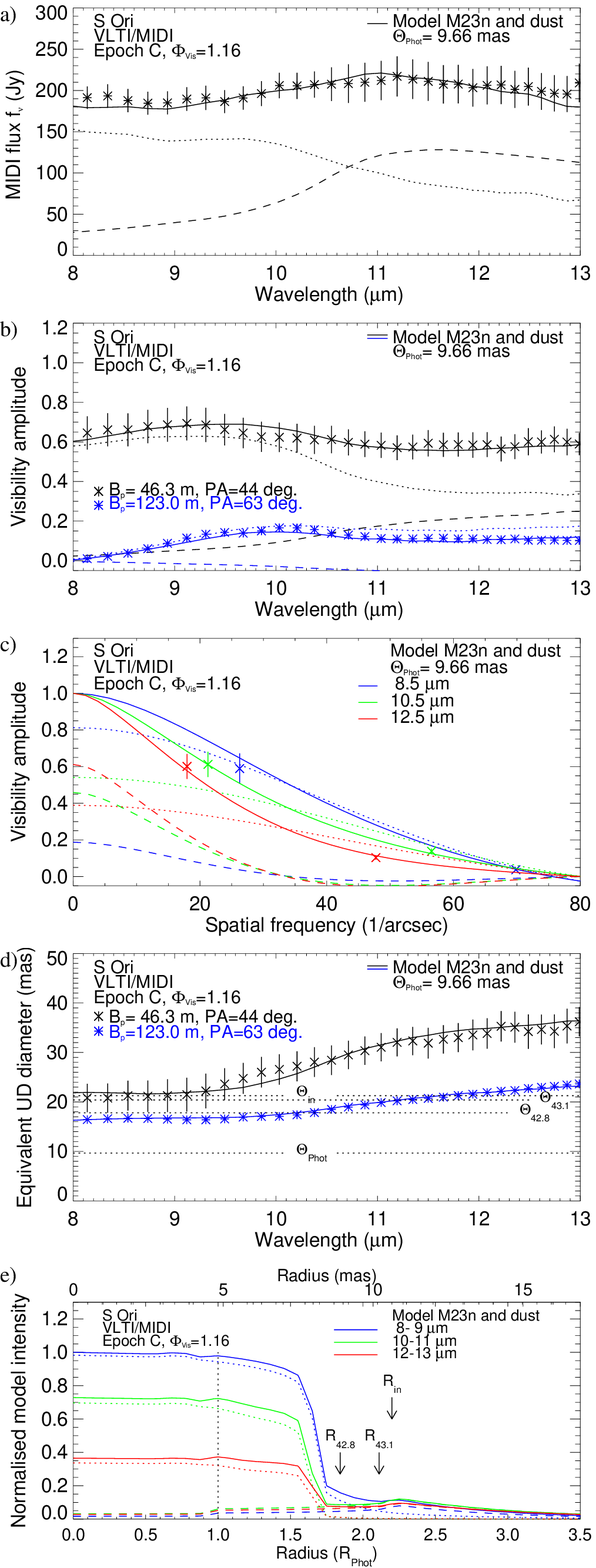}}
\caption{Results of our VLTI/MIDI 8--13\,$\mu$m interferometry
of S~Ori at epoch C, stellar phase 1.16. 
As Fig.~\protect\ref{fig:midiA}.
} 
\label{fig:midiC}
\end{figure}
\begin{figure}
\centering
\resizebox{1.0\hsize}{!}{\includegraphics{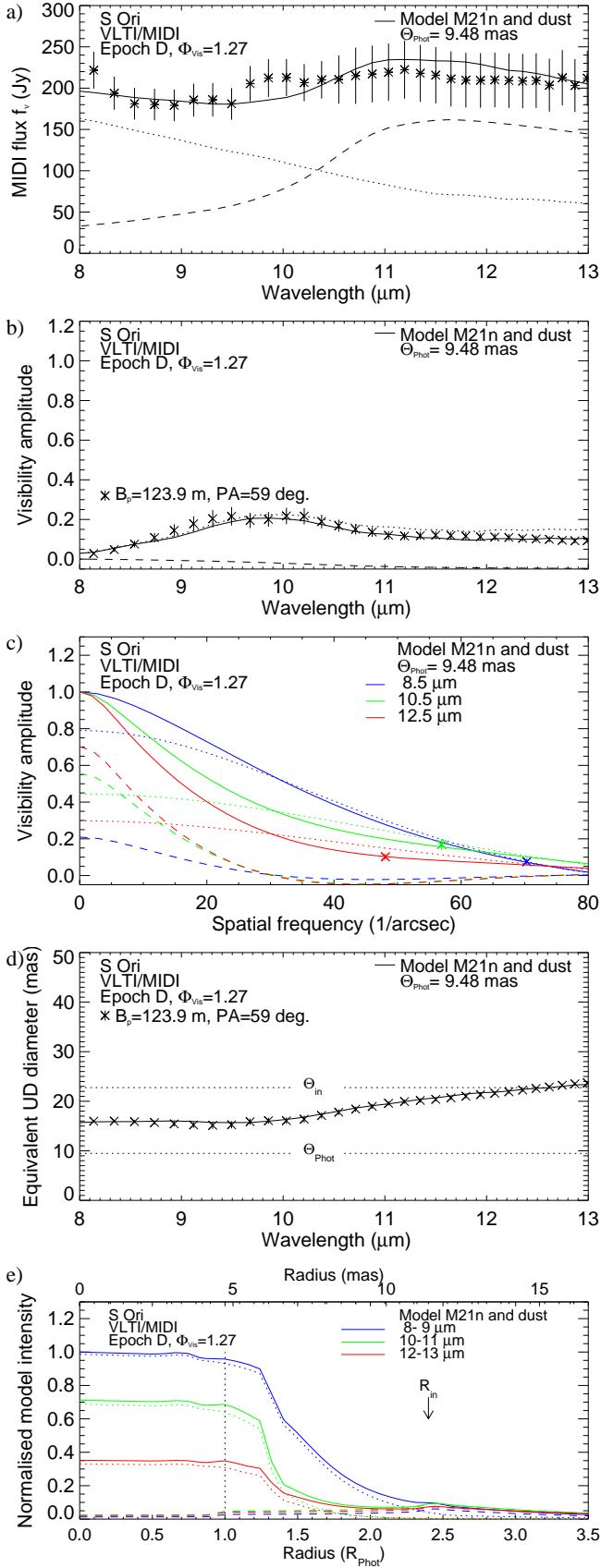}}
\caption{Results of our VLTI/MIDI 8--13\,$\mu$m interferometry
of S~Ori at epoch D, stellar phase 1.27. 
As Fig.~\protect\ref{fig:midiA}.
} 
\label{fig:midiD}
\end{figure}
\subsubsection{Dust-shell model for S~Ori}
Except for the dust-enshrouded very-long-period
Mira \object{IW~Hya} (Jeong et al. \cite{jeong03}),
models that self-consistently combine detailed dynamic 
atmosphere models {\it and} the formation of 
dust shells are currently not readily available for oxygen-rich 
AGB stars. However, the first attempts in this direction
have just been made (Ireland \& Scholz \cite{ireland06}, 
Woitke \cite{woitke06}). 

Dust shells around AGB stars have often been modeled using
ad-hoc radiative transfer calculations (see; for instance; 
Danchi et al. \cite{danchi94}; Lorenz-Martins \& Pompeia 
\cite{lorenz00}; Ohnaka et al. 
\cite{ohnaka05,ohnaka06a}), and we followed
this approach as the only currently available option
for describing S~Ori's dust shell.
We employed the Monte Carlo radiative transfer 
code {\tt mcsim\_mpi} (Ohnaka et al. \cite{ohnaka06a}) to 
calculate the dust shell model of S~Ori. For the central 
radiation source of the radiative transfer model, we used 
the spectral energy distribution (SED) of the respective 
M model atmosphere for the wavelength range 
from 0.32\,$\mu$m to 23\,$\mu$m and extended it by a 
blackbody approximation for longer wavelengths. 
Dust temperatures were calculated in radiative equilibrium.

Our MIDI spectra of S~Ori do not exhibit any prominent silicate 
feature, indicating that silicate grains are not a major 
constituent of the dust shell. Lorenz-Martins \& 
Pompeia (\cite{lorenz00}) have investigated the envelopes 
of 31 oxygen-rich AGB stars using the IRAS LRS spectra. They 
classify the objects into three groups, whose IRAS LRS spectra 
can be modeled using either silicate grains, Al$_2$O$_3$ 
grains, or a mix thereof. S~Ori is part of their study 
and belongs to the group of objects that can be modeled well
with Al$_2$O$_3$ grains alone. They show that a blend of 
alumina  Al$_2$O$_3$ grains (Koike et al. \cite{koike95}) 
for $\lambda<8\mu$m and porous amorphous  Al$_2$O$_3$ grains 
(Begemann et al. \cite{begemann97}) for $\lambda>8\mu$m gives 
the best agreement with the IRAS spectra, so we followed this 
approach (as in Ohnaka et al. \cite{ohnaka05} for the case of 
RR~Sco). Likewise, we considered small amounts of warm silicates
(Ossenkopf et al. \cite{ossenkopf92}) following the same works.
The grain size was set to 0.1\,$\mu$m for both Al$_2$O$_3$ and silicates.
We used the same inner boundary radius and density distribution 
for both dust species. The amount of dust of each species was 
characterized by the optical depths at a wavelength of 
0.55\,$\mu$m, denoted $\tau_V$(Al$_2$O$_3$) and $\tau_V$(silicate).
The density distribution $\rho(r)$ was described by a single 
power law with index $p$ ($\rho(r)\propto r^{-p}$).
We used the inner boundary radii in units of the 
stellar photospheric radius R$_\mathrm{Phot}$. 
All radiative transfer models of the dust shell
were calculated with an outer radius, so that the shell 
thickness is $R_\mathrm{out}/R_\mathrm{in}=1000$. 

\subsubsection{The main model parameters}
The stellar atmospheric structure including the effects from
atmospheric molecular layers is fully described by the choice of 
the specific model of the M series.
The M series consists of 20 models for different phase and cycle 
combinations as detailed in Ireland et al. (\cite{ireland04b}).
Here, we used a grid of the six M models of cycle 2 of the series,
which are M20 (model visual phase 0.05), M21n (0.10), 
M22 (0.25), M23n (0.30), M24n (0.40), and M25 (0.50). These 
models cover the range of our observations at phases 0.16 
(epoch C), 0.27 (epoch D), 0.42 (epoch A), and 0.55 (epoch D).
The absolute model phases with respect to observations
are uncertain by at least 0.1, with relative uncertainties 
of 0.01 to 0.02 between different models of the series 
(Ireland et al. \cite{ireland04b}). Likewise, the 
absolute assignment of visual phases to our epochs of 
observation is uncertain by about 0.1 
(see Sect.~\ref{sec:midiobs}). 
For each of these 6 atmosphere models, whose SEDs are used as 
central source, we calculated a grid of radiative transfer models 
of the dust shell. These grids included all combinations of optical depths 
$\tau_V$(Al$_2$O$_3$)=1.0, 1.5, 2.0, 2.5, 3.0; 
$\tau_V$(silicate)=0\%, 10\% of $\tau_V$(Al$_2$O$_3$); inner boundary radii
$R_\mathrm{in}/R_\mathrm{Phot}$=1.8, 2.0, 2.2, 2.4, 2.6;
and gradients $p$=2.0, 2.5, 3.0, 3.5, 4.0. 

Any overall model was thus defined with this set of
five independent parameters: (1) choice of the dust-free M 
model, (2) $\tau_V$(Al$_2$O$_3$), (3) $\tau_V$(silicate), 
(4) $R_\mathrm{in}$, (5) $p$. We used the well-defined 
continuum photospheric radius at $\lambda=$1.04\,$\mu$m $R_{1.04}$ 
of the respective dust-free M model to characterize the 
photospheric radius $R_\mathrm{Phot}$; i.e. we set
$R_\mathrm{Phot}=R_{1.04}$.
Any overall model is rigid; i.e. the distance of any model layer 
from $R_\mathrm{Phot}$ is fixed with the parameters given above.
The corresponding angular diameter $\Theta_\mathrm{Phot}$ was 
treated as the only fit parameter when comparing any such 
overall model to our MIDI data.
\subsubsection{Synthetic visibility and flux values}
\label{sec:synthetic}
\paragraph{Stellar contribution}
Monochromatic CLVs of the 
M models
were calculated at 3000 equally-spaced frequencies for the wavelength range  
8\,$\mu$m\,$\le \lambda \le 13$\,$\mu$m
as described in Ohnaka et al. (\cite{ohnaka06b}),
covering radii $r$ between 0 and $R^\mathrm{star}_\mathrm{out}=5\times R_p$.
These CLVs were calculated with H$_2$O and SiO (28SiO, 29SiO, and 30SiO) lines
included.  The line lists used for the calculations were the same as those
used for modeling RR~Sco by Ohnaka et al. (\cite{ohnaka05}); and 
as in the case of RR~Sco, we adopted silicon isotope ratios of 
28Si/29Si = 15 and 28Si/30Si = 20.
The H$_2$O and SiO line opacities were computed assuming local
thermodynamical equilibrium (LTE) and a Voigt profile with a
micro-turbulent velocity of 3\,km\,s$^{-1}$.  

The visibility values based on the stellar contribution were 
calculated for each spectral channel $c$ of MIDI as
\begin{equation}
V^\mathrm{star}_c = \frac{
\int_{\nu_\mathrm{min}(c)}^{\nu_\mathrm{max}(c)}\int_0^{R^\mathrm{star}_\mathrm{out}}\ I^\mathrm{star}_\nu(r)\ e^{-\tau(\nu)}\ J_0(\pi\ \Theta^\mathrm{star}_\mathrm{out}\ (B/\lambda))\ r\ dr\ d\nu
}
{
\int_{\nu_\mathrm{min}(c)}^{\nu_\mathrm{max}(c)}\int_0^{R^\mathrm{star}_\mathrm{out}}\ I^\mathrm{star}_\nu(r)\ e^{-\tau(\nu)} r\ dr\ d\nu.
}
\end{equation}
Here, $B$ is the projected baseline length used for the 
observation. Note that this computation is valid for the 
case of the {\tt EWS} data reduction, which coherently integrates 
the fringe signal to directly obtain the visibility modulus. 
For the {\tt MIA} analysis, which is a power spectrum analysis 
integrating the squared visibility modulus, the integration 
over frequency would need to be performed
over the squared monochromatic visibility values 
(cf. Wittkowski et al. \cite{wittkowski06}). 
The observed stellar flux contribution $f^\mathrm{star}_c$ 
was computed for each of MIDI's spectral channels as 
\begin{equation}
f^\mathrm{star}_c = 
2\ \pi \frac{\Theta^\mathrm{star}_\mathrm{out}[\arcsec]/2}{206265}\int_{\nu_\mathrm{min}(c)}^{\nu_\mathrm{max}(c)}\int_0^{R^\mathrm{star}_\mathrm{out}}\ I^\mathrm{star}_\nu(r)\ e^{-\tau(\nu)} r\ dr\ d\nu.
\end{equation}  
The spectral channels of MIDI were assumed to have a rectangular 
sensitivity function between frequencies $\nu_\mathrm{min}$ and  
$\nu_\mathrm{max}$. 
The term $e^{-\tau(\nu)}$ describes the
extinction by the dust shell. The optical depth $\tau(\nu)$ was 
calculated based on tabulated absorption and scattering 
coefficients $Q_\mathrm{abs}$ and $Q_\mathrm{sca}$ of the two 
dust species $i$ as
\begin{equation}
\tau(\nu)=\sum_i\ \tau^i(\nu_0)\ \frac{Q_\mathrm{abs}^i(\nu)+Q_\mathrm{sca}^i(\nu)}
{Q_\mathrm{abs}^i(\nu_0)+Q_\mathrm{sca}^i(\nu_0)},
\end{equation}
where we used $\nu_0$ corresponding to $\lambda_0=0.55\mu$m. 
\paragraph{Dust contribution}
\label{sec:dustclv}
Using the dust temperatures and monochromatic mean intensities computed
by a Monte Carlo run with {\tt mcsim\_mpi} for a given set of dust shell
parameters, we calculated monochromatic CLVs of the dust shell (excluding the
contribution of the central star) with the ray-tracing method as 
described in Ohnaka et al. (\cite{ohnaka06a,ohnaka07}).
The radii $r$ covered the range between 0 and 25\,$R_\mathrm{Phot}$,
and we denote the outermost model radius, where we set the 
intensity to 0, 
by $R^\mathrm{dust}_\mathrm{out}=25\times R_\mathrm{Phot}$.
The outer radius of 25\,$R_\mathrm{Phot}$ was chosen because
the MIDI interferometric field of view of $\sim$\,250\,mas
corresponds to 25 times the expected photospheric 
angular diameter of S~Ori of $\sim$\,10\,mas. 
We verified that changing the outer model radius 
from 25\,$R_\mathrm{Phot}$ to 50\,$R_\mathrm{Phot}$ changed the 
diameter results by less than 0.3\%. This indicates that the 
contribution from dust to the overall CLV outside of 25 photospheric 
radii can be neglected and that the exact choice of the
outermost model radius is not critical for our calculation.
Visibility ($V^\mathrm{dust}_c$) and flux ($f^\mathrm{dust}_c$) 
values for each spectral channel of MIDI were computed in the 
same way as for the dust-free model atmosphere, but 
without the extinction term $e^{-\tau(\nu)}$.
\paragraph{Comparison of visibility and flux values to our data}
Finally, the flux and visibility values of the overall model
were computed as
\begin{equation}
f^\mathrm{total}_c=f^\mathrm{star}_c + f^\mathrm{dust}_c
\end{equation}
and
\begin{equation}
V^\mathrm{total}_c= \frac{f^\mathrm{star}_c}{f^\mathrm{total}_c}\ V^\mathrm{star}_c +
\frac{f^\mathrm{dust}_c}{f^\mathrm{total}_c}\ V^\mathrm{dust}_c.
\end{equation}

This simple addition of the CLVs of the dust-free model atmosphere of the
M series and of the pure dust-shell (i.e. after subtraction of the 
contribution by the attenuated input radiation) is fully valid as 
long as the stellar atmosphere and dust shell are spatially separated. 
In fact, there may be some overlap of the extended low-intensity 
atmospheric molecular layers of the dust-free atmosphere model and 
the dust shell in our overall models. However, 
we verified that the inner boundary radius of the dust shell 
is clearly larger than that of the layer at which the 
atmosphere becomes optically thick at mid-infrared wavelengths.
Ireland \& Scholz (\cite{ireland06}) find for similar conditions
that the gas temperatures for a dust-free M model and an M model 
that includes re-heating of gas by dust, differ by less than 50\,K for 
radii up to about 3 parent-star radii.

For any given overall model separately, we optimized the angular 
diameter $\Theta_\mathrm{Phot}$ using a standard least-square fit that 
minimizes the $\chi^2$ value between synthetic and observed 
visibility {\it and} flux values. Hereby, $\Theta_\mathrm{Phot}$
was treated as the only free parameter. The resulting reduced
$\chi^2_\nu$ values among our full grid of different overall 
models were compared in order to find the best-fitting set of 
model parameters. This analysis was separately performed for 
each of our epochs A--D. 
\subsection{MIDI model results}
\label{sec:midimodelresults}
\begin{table}
\caption{Fit results of our MIDI measurements of S~Ori.}
\label{tab:midiresults}
\begin{tabular}{lrrrr}
\hline\hline
Epoch ($\Phi_\mathrm{Vis}$) & A (0.42) & B (0.55) &  C (1.16) & D (1.27) \\\hline
Best model  & M22 & M24n & M23n & M21n  \\
$\Phi_\mathrm{Model}$              & 0.25 & 0.40 & 0.30 & 0.10\\
$\phi_\mathrm{Model}-\phi_\mathrm{Vis}$ & -0.17 & -0.15 & +0.14 & -0.17 \\
Best $R_\mathrm{in}$/$R_\mathrm{Phot}$         & 1.8   &   2.0 & 2.2   &  2.4 \\
Best $\tau_V$(Al$_2$O$_3$)  &   2.5  & 2.5 & 1.5 & 1.5 \\
Best $\tau_V$(Silicate)  & 0 & 0 & 0 & 0 \\
Best $p$                 & 3.5 & 3.5 & 3.0 & 2.5 \\ 
$\Theta_\mathrm{Phot}$ [mas]    & 9.00  & 7.92 & 9.66  & 9.48 \\
$\chi^2_\nu$             & 0.16  & 0.10 & 0.43  & 0.35 \\
$n$                      & 62 & 186 & 124 & 62 \\
$\sigma_\mathrm{formal} $ [mas] &  0.10  &  0.05  & 0.02 & 0.07  \\
$\sigma_\mathrm{model}$ [mas]   &  0.30  &  0.09  & 0.06 & 0.34  \\\hline
\end{tabular}
\end{table}
\begin{table}
\caption{Reduced $\chi^2_\nu$ values obtained for our grid of 
overall models. }
\label{tab:chi2}
\begin{tabular}{lr|rrrr}
Epoch   &       &    A     &    B     &    C     &    D     \\\hline 
M model &  M20  &   0.27   &  0.16    &   0.52   &  0.38       \\
        &  M21n &   0.26   &  0.14    &   0.53   & \fbox{0.35} \\
        &  M22  & \fbox{0.16} & 0.17     &  0.81   &  0.51    \\
        &  M23n &    0.17  &  0.12    & \fbox{0.43} &  0.61    \\
        &  M24n &    0.20  & \fbox{0.10} &   0.66   &  1.04    \\
        &  M25n &    0.19  &  0.17    &   0.53   &  0.58    \\\hline

$R_\mathrm{in}$/$R_\mathrm{Phot}$     
        & 1.8   & \fbox{0.16} & 0.10     &  0.56    &  0.54    \\
        & 2.0   &   0.19   & \fbox{0.10} &  0.52    &  0.46    \\
        & 2.2   &   0.21   & 0.11     & \fbox{0.43} &  0.39    \\
        & 2.4   &   0.24   & 0.12     &  0.50    & \fbox{0.35} \\
        & 2.6   &   0.29   & 0.18     &  0.57    & 0.38     \\\hline
                                
$\tau_V$(Al$_2$O$_3$) 
        & 1.0   &  /       & /        & 0.77     &  0.48    \\
        & 1.5   &  0.18    & 0.11     & \fbox{0.43} & \fbox{0.35} \\
        & 2.0   &  0.17    & 0.11     & 0.56     & 0.54     \\
        & 2.5   & \fbox{0.16} & \fbox{0.10} & 0.80     & 0.71     \\
        & 3.0   &  0.18    & 0.11     &  /        &    /      \\\hline
                                            
 $\tau_V$(Silicate)/           
        & 0\%   & \fbox{0.16} & \fbox{0.10} & \fbox{0.43} & \fbox{0.35}     \\
$\tau_V$(Al$_2$O$_3$) 
        &10\%  & 0.19      & 0.12     & 0.52     & 0.38 \\\hline

$p$     & 2.0   & 0.31     &  0.20    &  0.87    & 0.38     \\
        & 2.5   & 0.20     &  0.14    &  0.54    & \fbox{0.35} \\
        & 3.0   & 0.17     & 0.10     & \fbox{0.43} & 0.39     \\
        & 3.5   & \fbox{0.16} & \fbox{0.10} &  0.50    & 0.51     \\
        & 4.0   & 0.20     & 0.12     &  /        &  /           \\\hline
\end{tabular}
\end{table}
Table~\ref{tab:midiresults} lists for each of our MIDI epochs 
the parameter set of the best-fitting overall model of our grid.
Listed are for each of our 4 epochs the best-fitting 
atmosphere model of the M series; its model phase; 
the difference between observed and model phase; 
the best-fitting parameters of the dust shell 
($\tau_V$(Al$_2$O$_3$), $\tau_V$(Silicate), 
$\tau_V$(Silicate), $p$); the best-fitting angular diameter 
$\Theta_\mathrm{Phot}$; the reduced $\chi^2_\nu$ value; 
the number of data points used ($n$); and the formal and model errors.
The model errors are 
derived as the standard deviations of $\Theta_\mathrm{Phot}$ values 
based on different overall models of our grid that are 
consistent at the 1\,$\sigma$ level based on application of
the $F$-test. The model errors of epochs A and D are larger than 
those of epochs B and C, which can be explained by the use of 
only one baseline configuration at these epochs. 
Table~\ref{tab:chi2} lists for each epoch and for each grid parameter the
best $\chi^2_\nu$ value that could be obtained by any combination 
of all other parameters. The best $\chi^2_\nu$ 
value for each epoch is marked by boxes.

The synthetic visibility
and flux values corresponding to the best-fitting overall 
models are shown in Figs.~\ref{fig:midiA}-\ref{fig:midiD} 
in comparison to the measured values. Panels (d), which show 
the equivalent UD diameter values, also 
indicate the photospheric angular diameter
$\Theta_\mathrm{Phot}$ and the inner dust shell 
boundary $\Theta_\mathrm{in}$ derived from the model fit. Panels (e) show 
the intensity profiles of the overall models that fit 
our MIDI data best for three bandpasses at 8--9\,$\mu$m, 
10--11\,$\mu$m, and 12--13\,$\mu$m. These panels have a 
bottom and a top abscissa indicating (bottom) the radius as 
a function of the model photospheric radius $R_\mathrm{Phot}$ 
and (top) the corresponding angular scale using the best-fitting 
angular diameter $\Theta_\mathrm{Phot}$.
Panels (e) also indicate the inner dust shell 
radius $R_\mathrm{in}$, as well as the mean SiO maser 
ring radii $R_{42.8}$ and $R_{43.1}$, which are derived 
in Sect.~\ref{sec:vlba}. A two-dimensional pseudo-color
image of the mid-infrared model intensity showing the molecular
layers and the dust shell is displayed 
in Fig.~\ref{fig:allepochs}. 
The photospheric disk is represented by a light blue color.
Overlaid are the images of the SiO 
maser radiation described in Sect.~\ref{sec:vlba}.
\paragraph{The photospheric diameter $\Theta_\mathrm{Phot}$}
The obtained photospheric diameter values $\Theta_\mathrm{Phot}$ 
at phases 0.42, 0.55, 1.16, and 1.27 are 9.0\,mas $\pm$ 0.3\,mas, 
7.9\,mas $\pm$ 0.1\,mas, 9.6\,mas $\pm$ 0.1\,mas, 
and 9.5\,mas $\pm$ 0.4\,mas, respectively. Here, the error bars
include the formal, as well as the model, errors 
from Table~\ref{tab:midiresults}.
Our measurements indicate significant changes of the 
photospheric diameter $\Theta_\mathrm{Phot}$ as a function of 
stellar phase. 
A detailed comparison of $\Theta_\mathrm{Phot}$ to the 
inner dust shell radius and the SiO maser ring radii follows in 
Sect.~\ref{sec:comparison}.

In BW05, we estimated S~Ori's photospheric 
angular diameter at phase 0.73 to $\sim$\,9.2\,mas, based 
on our $K$-band UD diameter measurements and 
using correction factors from UD to photospheric 
diameters from Ireland et al. (\cite{ireland04a,ireland04b}). 
This value is consistent with the photospheric diameters 
derived here, as the estimate of 9.2\,mas at phase 0.73 
from BW05 lies as expected in between the values 
of 7.9\,mas at phase 0.54 and 9.6\,mas at phase 1.16 derived 
in the present paper.

The $K$-band UD diameter of 10.54\,mas $\pm$ 0.68\,mas 
by van Belle et al. (\cite{vanbelle96}) was obtained at 
phase 0.56. At phases close to this value, the M series 
results in a ratio between the $K$-band UD diameter 
and the photospheric diameter of $\sim$\,1.35 (Fig.~7 
of Ireland et al. \cite{ireland04b}). With this correction, 
the photospheric diameter corresponding to this measurement 
at phase 0.56 would be 7.8\,mas $\pm$ 0.5\,mas, 
and would thus be fully consistent with our value 
of 7.9\,mas $\pm$ 0.1\,mas at the same phase (Epoch B at 
phase 0.54).

Millan-Gabet et al. (\cite{millan05}) obtained a $K$-band 
UD diameter of 9.6\,mas $\pm$ 0.2\,mas at phase 0.1.
At this phase, the ratio between $K$-band UD diameter
and photospheric diameter is predicted to be close to
unity (Ireland et al. \cite{ireland04b}). 
This measurement thus also agrees well with our 
photospheric diameter of 9.6\,mas $\pm$ 0.4\,mas at 
phase 1.16 (epoch C).

The good agreement of the photospheric angular diameters derived
from our MIDI measurements in the present paper with 
previous near-infrared diameter measurements
increases confidence in the validity
of our modeling using the M model series complemented by
a radiative transfer model of the dust shell.
\paragraph{The choice of the M model}
The intensity profile of the extended atmosphere, and thereby
the choice of the best-fitting model out of the M series, 
is mainly constrained by the shape of the visibility functions,
in particular for longer baseline lengths, in the wavelength 
range $\sim$\,8--9\,$\mu$m, where the contribution by the dust 
shell is relatively low.

The differences in visual stellar phase of the best-fitting 
M model compared to the visual phase at the epochs of 
observation are -0.17, -0.15, +0.14, and -0.17, respectively. 
These differences may be due to the uncertainties of the 
absolute assignments of visual phases
to the model series as well as to the dates of observation, 
which are both on the order of 0.1. The relative
phase values are expected to agree better, which is 
true for epochs A, B, and D. Differences between these phases
might also occur owing to the different radius and pulsation period
of the model series' parent star compared to S~Ori.
These parameters might have an effect on the extension of
the atmosphere, so that the best-fitting extension of the
atmosphere might belong to a model at a different visual phase.
An effect such as this might explain the clearly different 
phase difference between observation and model at epoch C 
compared to the three other epochs. 

The good agreement of the shape of the synthetic visibility 
function with our observations at the wavelength range 
from 8\,$\mu$m to 9\,$\mu$m and the simultaneous agreement 
with the measured flux level again confirms that the 
M model series is usable for S~Ori.

As an indication of the extension of the molecular layers of the 
best-fitting M models, the 50\%/10\% intensity radii of the 
$N$-band (8-13\,$\mu$m) model CLVs are 1.5/1.6, 1.8/1.9, 1.6/1,7,
and 1.4/1.9 photospheric radii $R_\mathrm{Phot}$ for epochs A-D, i.e. for 
models M22, M24n, M23n, and M21n, respectively.
\paragraph{The dust shell chemistry}
The ratio of the number of Al$_2$O$_3$ grains to that of 
silicate grains is mostly constrained by the shape of the 
flux and visibility spectra at the silicate feature close 
to 10\,$\mu$m. We obtain best fits to our S~Ori
data with a dust shell only consisting of Al$_2$O$_3$ grains. 
The addition of small amounts of silicate grains does not 
improve the fits for any of our epochs. While our grid in 
Table~\ref{tab:chi2} only includes silicate grains with the 
same inner boundary radius as for the Al$_2$O$_3$ grains,
we also compared to our data the best-fitting models 
from Table~\ref{tab:midiresults} with silicate grains added at 
larger inner boundary radii of 4 and 6 photospheric
radii. These models lead to clearly increased $\chi^2_\nu$
values as well. In summary, our data indicate that silicon is 
not bound in dust, neither at very close distances of 
$\sim$\,2\,R$_\mathrm{Phot}$ nor at larger distances of 
4-6\,R$_\mathrm{Phot}$.
A dust shell only consisting of Al$_2$O$_3$ grains is 
consistent with the modeling of the
IRAS LRS spectra of S~Ori by Lorenz-Martins \& Pompeia (\cite{lorenz00}). 
\paragraph{The radial density profile of the dust shell}
\label{sec:dust}
The total optical depth of the dust shell is mainly constrained
by the height and shape of the MIDI total flux spectrum. The 
radial density profile described by the inner boundary radius and the density 
gradient is mostly constrained by the shape of the visibility function 
versus wavelength. Here, the values at shorter wavelengths, 
probing warmer dust, constrain the inner boundary radius best.
The density gradient is mostly constrained by the visibility
values at longer wavelengths and shorter baseline lengths, 
with which the dust shell is less resolved.  

We obtain best-fitting inner boundary radii of 
1.8, 2.0, 2.2, and 2.4 photospheric radii $R_\mathrm{Phot}$
at our phases 0.42, 0.55, 1.16, and 1.27, respectively.
These radii are well-constrained and do not significantly
change for variations in other model parameters.
We estimate the uncertainty of this parameter to about 
$0.2\times R_\mathrm{Phot}$. The formation of dust consisting
of Al$_2$O$_3$ grains at these short distances from the
stellar surface is consistent with the empirical
results by Lorenz-Martins \& Pompeia (\cite{lorenz00}),
as well as with the recent theoretical calculations 
by Ireland \& Scholz (\cite{ireland06}) 
and Woitke (\cite{woitke06}).

Best-fitting density gradients $p$ at the same phases
are 3.5, 3.5, 3.0, and 2.5, respectively, and 
best-fitting optical depths $\tau_V$ are 2.5, 2.5,
1.5, and 1.5. 
Taking the inner boundary radii, the density gradients, 
and the optical depth values together, our fits indicate
more compact dust shells with larger optical depth near
the stellar minimum and more extended dust shells with lower
optical depth after stellar maximum.
\paragraph{Wind models}
\begin{figure}
\centering
\resizebox{1\hsize}{!}{\includegraphics{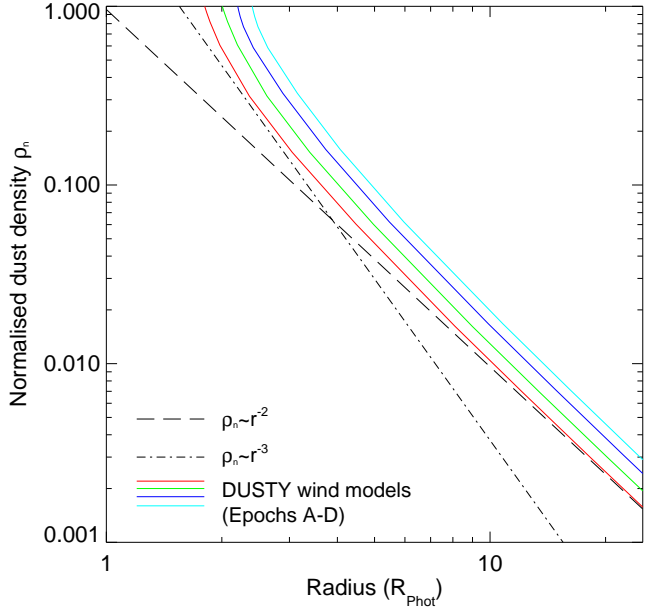}}
\caption{Density profiles of radiatively driven wind models calculated 
with the {\tt DUSTY} code (Ivezi{\'c} \& Elitzur \cite{dusty1}, 
Ivezi{\'c} et al. \cite{dusty2}) for input parameters from 
Table~\ref{tab:midiresults}, together with $\rho\propto r^{-2}$ 
and $\rho\propto r^{-3}$ curves.}
\label{fig:densplot}
\end{figure}
We used the radiation transport code {\tt DUSTY} 
(Ivezi{\'c} \& Elitzur \cite{dusty1}, Ivezi{\'c} et al. \cite{dusty2})
with its option of modeling radiatively driven winds to compute the density
structure of S~Ori's envelope that would result from a wind model.
We computed wind models for each of our epochs A to D using
a blackbody with effective temperature of the best-fitting M model 
as the central radiation source and with the $R_\mathrm{in}$ 
and $\tau_V$ values
from Table~\ref{tab:midiresults}. We used the same optical properties of
Al$_2$O$_3$ as above, and we set $R_\mathrm{out}/R_\mathrm{in}$ to $10^3$. 
Figure~\ref{fig:densplot} shows the resulting normalized density profiles
of the wind models together with the $\rho\propto r^{-2}$ 
and $\rho\propto r^{-3}$ 
curves. The wind models result in density structures $\rho\propto r^{-2}$
for $r\ga$\,10\,$R_\mathrm{Phot}$. Closer to the 
photosphere, the gradients of the wind density profiles become steeper
than $\rho\propto r^{-3}$. As our MIDI 
visibility measurements are mostly sensitive to the inner dust region, 
the relatively high density gradients given
in Table~\ref{tab:midiresults} could be explained
by such a wind model. The differences among the empirically found density
gradients for the different epochs could be explained by different mean
distances from which most mid-infrared photons escape.
The wind models shown correspond to mass-loss rates
between $\sim$\,2.4\,10$^{-6}$\,M$_\odot$/yr (Epoch C) and 
$\sim$\,3.5\,10$^{-6}$\,M$_\odot$/yr 
(Epoch B). The indication of more compact 
dust shells with larger optical depth near stellar minimum and more 
extended dust shells with lower optical depth after stellar maximum 
thus suggests a wind with increased mass loss and dust formation occurring
relatively close to the stellar surface near minimum visual phase and with
an expanded dust shell after maximum phase.
\paragraph{Symmetry of the molecular and dust shells}
Our MIDI data at epochs A and D are each obtained at one single position
angle on sky. Epoch B spans a range in position angles of 
$\sim$\,40\,$\,\degr$ and epoch C of $\sim$\,20\,$\,\degr$. Within these
probed ranges, our S~Ori data can be described well by spherically symmetric
distributions of molecular layers and dust shells without the need to 
introduce deviations from circular symmetry to our modeling.
However, we cannot rule out asymmetries at wider ranges of position
angles, in which case the resulting parameters in Table~\ref{tab:midiresults}
are only valid along the position angles on the sky given in 
Table~\ref{tab:vltiobs}.
Note that the distributions of the SiO maser spots toward S~Ori shown in 
BW05, as well as below in Sect.~\ref{sec:vlba}, do not indicate 
a globally asymmetric gas distribution.  
\section{VLBA measurements}
\label{sec:vlba}
\label{sec:vlbaobs}
Contemporaneous with the VLTI/MIDI observations, we observed the
$v=1, J=1-0$ (43.1\,GHz) and $v=2, J=1-0$ (42.8\,GHz) SiO maser
emission associated with S~Ori ($\alpha = 05^h 29^m 00^{s}.9,
\delta = $-$04^{\circ} 41' 32''.7$, J2000) using the 10 stations
of the VLBA, which is operated by the National Radio Astronomy
Observatory (NRAO). Table~\ref{tab:vlbaobs} lists the details of our
SiO maser observations recorded over three epochs. Listed are the epoch,
date, time, Mean Julian Date (MJD) of the start of observation,  
experiment duration, number of antennas, and the 
visual phase $\Phi_\mathrm{vis}$.
The first two epochs occurred near stellar minimum and were
spaced 42 days apart or approximately 10\% of the stellar period.
The third epoch occurred roughly 9 months later just after maximum
on the next stellar cycle.
Data were
recorded in a similar manner for all three epochs at reference
frequencies of 43.122080\,GHz and 42.820587\,GHz for the $v=1$
and $v=2$ SiO transitions, respectively.  Right- and left-hand
circular polarization was used with 8-MHz (56.1\,km\,s$^{-1}$) bands
centered on the local standard-of-rest (LSR) velocity 
of 18.0\,km\,s$^{-1}$.
Interspersed with observations of S~Ori were 10-minute
scans of extragalactic continuum sources (0423$-$013 and 0359$+$509)
for the purpose of delay and bandpass calibration.
\begin{table}
\caption{VLBA observations of S~Ori.}
\label{tab:vlbaobs}
\begin{tabular}{llllrrl}
\hline\hline
Ep. & Date & Time & MJD & Dur. & No. &$\Phi_\mathrm{vis}$ \\
     &      & UTC  &-2.4E6 &    (hrs)  &   Ant. &               \\\hline
A   & 2005-01-17 & 02:36 & 53387.6 & 5.0  & 9   & 0.46 \\
B   & 2005-02-28 & 23:47 & 53430.5 & 2.6 & 10  & 0.56 \\
C   & 2005-11-05 & 07:23 & 53679.8 & 5.0 & 10 & 1.14 \\
\hline
\end{tabular}
\end{table}
For epoch A only 9 of the 10 stations of the VLBA 
were available for our observations.  Unfortunately, the one antenna 
that was unavailable was the station at Mauna Kea, HI.  This station 
when combined with the St. Croix, VI antenna provides the longest 
baseline  (greatest resolution) in the array. For this reason the 
resolution of the epoch A images is reduced from that of the other two 
epochs.
There was also a problem with the second epoch (epoch B) of 
observations in that roughly half of the data were accidentally lost 
after the correlation. Despite this loss, we were still able to use 
the remaining data to produce quality images of the masers. NRAO kindly 
scheduled a third epoch of observations (epoch C) to make up for the loss.
\subsection{VLBA data reduction and analysis}
\label{sec:vlba_reduction}
The data were correlated at the VLBA correlator operated by NRAO in 
Socorro, New Mexico.  Auto and cross-correlation spectra consisting 
of 256 channels with channel spacings 
of 31.25\,kHz ($\sim$\,0.2\,km\,s$^{-1}$) 
were produced by the correlator. Calibration was performed using the 
Astronomical Image Processing System (AIPS) maintained by NRAO.  The 
data were calibrated in a manner similar to what was performed in BW05.
For each epoch, residual delays due to the instrumentation were corrected 
by performing a fringe fit on the continuum calibrator scans. Residual group 
delays for each antenna were determined and applied to the spectral line 
data.  Variations in the residual delays ranged from 2-4\,ns resulting in 
phase errors of no more than 1.5-3$^{\circ}$ across the 8-MHz band.

The bandpass response was determined from scans on the continuum
calibrators and used to correct the target-source data.
The time-dependent gains of all antennas relative to a reference
antenna were determined by fitting a total-power spectrum (from the
reference antenna with the target source at a high elevation) to the
total power spectrum of each antenna. The absolute flux density
scale was established by scaling these gains by the system temperature and
gain of the reference antenna. Errors in the gain and pointing of
the reference antenna and atmospheric opacity variations contribute to
the error in the absolute amplitude calibration, 
which is accurate to about 15--20\%.

Residual fringe rates were obtained by fringe-fitting a strong reference 
feature in the spectrum of each maser transition.  For epoch A we used the
channel at velocity $V_{\rm LSR} = 12.3$\,km\,s$^{-1}$ for the $v=1$ 
transition and the channel at velocity $V_{\rm LSR} = 16.9$\,km\,s$^{-1}$ 
for the $v=2$ transition. For epoch B we used channels 
at $V_{\rm LSR} = 12.8$\,km\,s$^{-1}$ and $V_{\rm LSR} = 12.1$\,km\,s$^{-1}$ 
for the $v=1$ and $v=2$ transitions, respectively.  Finally, for epoch C 
we used the same velocity channel for both transitions 
$V_{\rm LSR} = 10.3$\,km\,s$^{-1}$.  The resulting fringe-rate solutions 
were applied to all channels in each spectrum. An iterative 
self-calibration and imaging procedure was then performed to map this 
reference channel for each transition. The resulting residual phase and
amplitude corrections from the reference channels at 42.8\,GHz
and 43.1\,GHz were applied to all channels in the respective
bands. This procedure was repeated for each transition at
each epoch.

In order to accurately compare the distributions of the two
maser transitions, it is desirable to determine a common
spatial reference point.  However, after the fringe-fitting
step to determine residual fringe-rates, all absolute position
information is lost for the VLBA data. To accomplish this
registration, we used the same technique employed in BW05 with
the residual fringe-rates, phases, and
amplitudes from one transition (e.g., $v=2$, 42.8\,GHz) used to
calibrate the other transition (e.g., $v=1$, 43.1\,GHz) in
order to determine an offset between the two phase centers.

\begin{figure*}
\centering
\resizebox{1.\hsize}{!}{\includegraphics{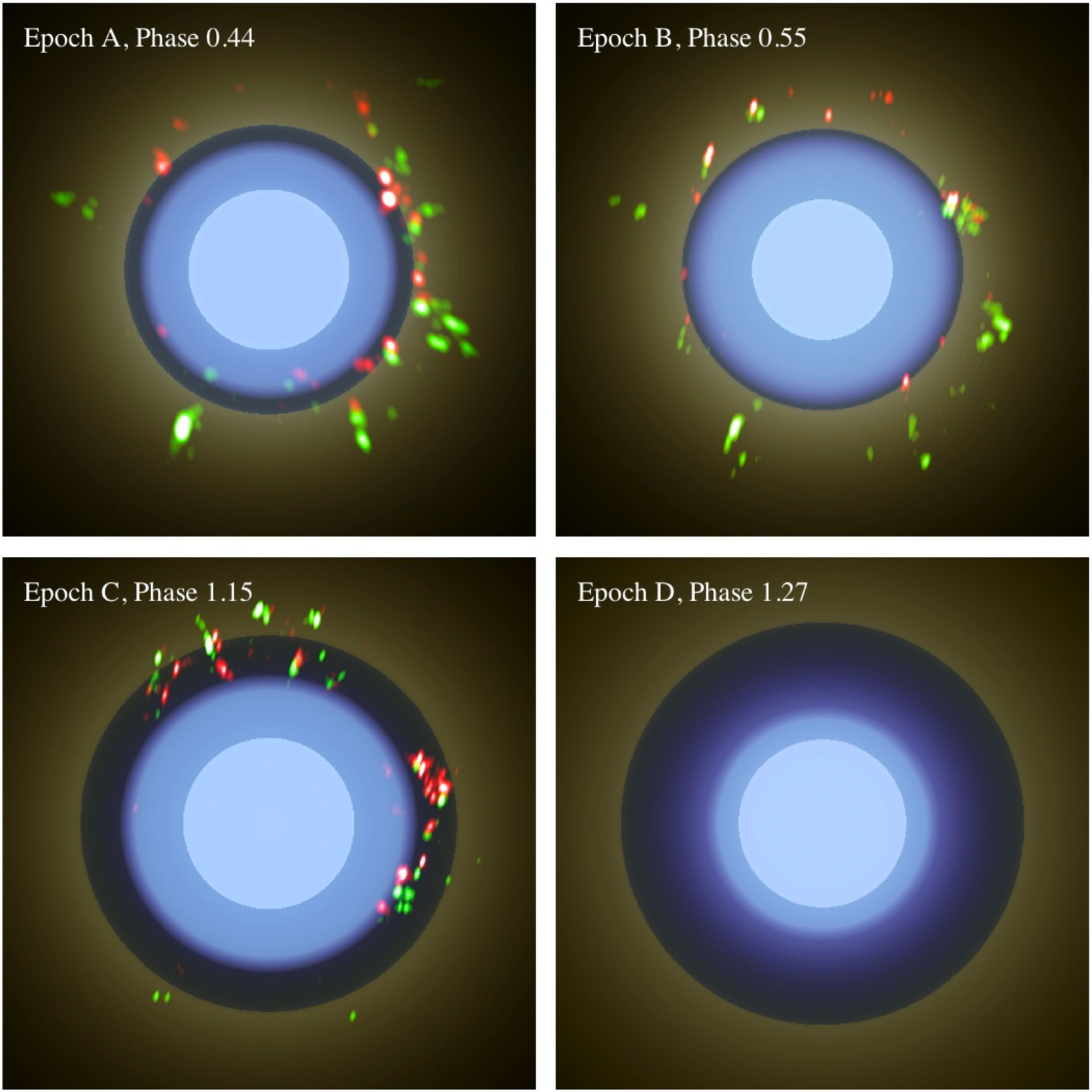}}
\caption{The (red) $v=2$, $J=1-0$ (42.8\,GHz) and 
(green) $v=1$, $J=1-0$ (43.1\,GHz) maser images overlaid onto
pseudo-color representations of the infrared intensity. 
The continuum photosphere (in fact mostly hidden behind the molecular 
atmosphere in the $N$-band) is enhanced to a light blue color. The 
darker blue shades represent our model intensity profile as in panels
(e) of Figs.~\ref{fig:midiA}--\ref{fig:midiD}, and the green shades
represent the location of the Al$_2$O$_3$ dust shell on top of the
low-intensity extended wings of the molecular atmosphere. The true location
of the star relative to the maser images is unknown. Here, we assume that the
center of the star coincides with the 
center of the maser spot distribution. 
Synthesized beam sizes 
for epoch A are $0.59\times 0.40$~mas and $0.60\times 0.37$~mas for
the $v=1$ and $v=2$ transitions, respectively. For epoch B these
sizes are $0.52\times 0.21$~mas and $0.51\times 0.20$~mas, and for epoch C 
$0.46\times 0.16$~mas and $0.44\times 0.16$~mas, respectively. Epoch D is
a MIDI-only epoch and SiO maser observations have not been obtained at 
this epoch. The size of each panel
is $30\times 30$\,mas, corresponding to $14.4\times 14.4$\,AU. }
\label{fig:allepochs}
\end{figure*}

The final image cubes of the SiO maser emission consisting
of $1024\times 1024$ pixels ($\sim$$51\times51$~mas) were generated
for each transition at each of the three epochs.
Images were produced for all spectral channels containing SiO maser 
emission in each transition at each epoch. Synthesized beam sizes 
used for epoch A were $0.59\times 0.40$~mas and $0.60\times 0.37$~mas for
the $v=1$ and $v=2$ transitions, respectively. For epoch B these
sizes were $0.52\times 0.21$~mas and $0.51\times 0.20$~mas. 
For epoch C the beam sizes used in the imaging were
$0.46\times 0.16$~mas and $0.44\times 0.16$~mas.
Note that the resolution was greatly reduced in epoch A due to the
loss of the Mauna Kea, HI VLBA antenna mentioned previously. The
resolution of epoch B is slightly worse than that of epoch C due to
the loss of half the data, thus reducing the $uv$ coverage of the
array. Using the procedure described above, the image cubes from the two
transitions were spatially aligned for each epoch.
Off-source {\em rms} noise estimates in the images ranged
from 5~mJy to 9~mJy for epoch A, 3~mJy to 7~mJy for epoch B,
and 7~mJy to 18~mJy for epoch C. Figure~\ref{fig:allepochs}
shows the total intensity images of the $v=1$, 43.1-GHz
and $v=2$, 42.8-GHz SiO maser emission toward S Ori
overlaid onto a pseudo-color representation of the mid-infrared
intensity profile obtained in Sect.~\ref{sec:vlti}.

The six resulting image cubes (2 transitions at 3 epochs) were
then analyzed to extract relevant maser parameters.
Two-dimensional Gaussian functions were fit to maser
emission above a cut-off flux density of 100\,mJy in each
spectral (velocity) channel using the AIPS task SAD.  This
fitting yielded velocities, flux densities, and positions in right
ascension and declination for all emission components identified
in the images.  Errors in right ascension and declination of
identified components were computed using the fitted source size
divided by twice the signal-to-noise ratio (s/n) in the image
and ranged from 1\,$\mu$as for features with high s/n, to 28\,$\mu$as
for features with lower s/n.

The remaining analysis of the maser component identifications was
performed outside of the AIPS package.  For comparison with the
total flux densities derived from the MIDI data, we summed the flux
densities and their errors for all fitted components in all channels
for each transition/epoch.  These integrated flux densities are
discussed further in Sect.~\ref{sec:comparison}.  
Since the $\sim$0.2\,km\,s$^{-1}$ channel spacing is sufficient to 
resolve the masers spectrally, features typically appear in multiple 
adjacent spectral channels.  Positions in right ascension and declination 
and center velocities for the masers were determined using a 
flux-density-squared weighted average for features identified in two or 
more adjacent channels with a spatial coincidence of 0.5\,mas. The flux 
assigned to the maser averages was the maximum single-channel flux density.  
The results from the image analysis and component averaging are described 
in the next section.
\subsection{VLBA results}
\label{sec:vlba_results}
\subsubsection{The spatial structure of the SiO masers}
\label{sec:vlba_spatial}
\begin{figure*}
\centering
\resizebox{0.325\hsize}{!}{\includegraphics{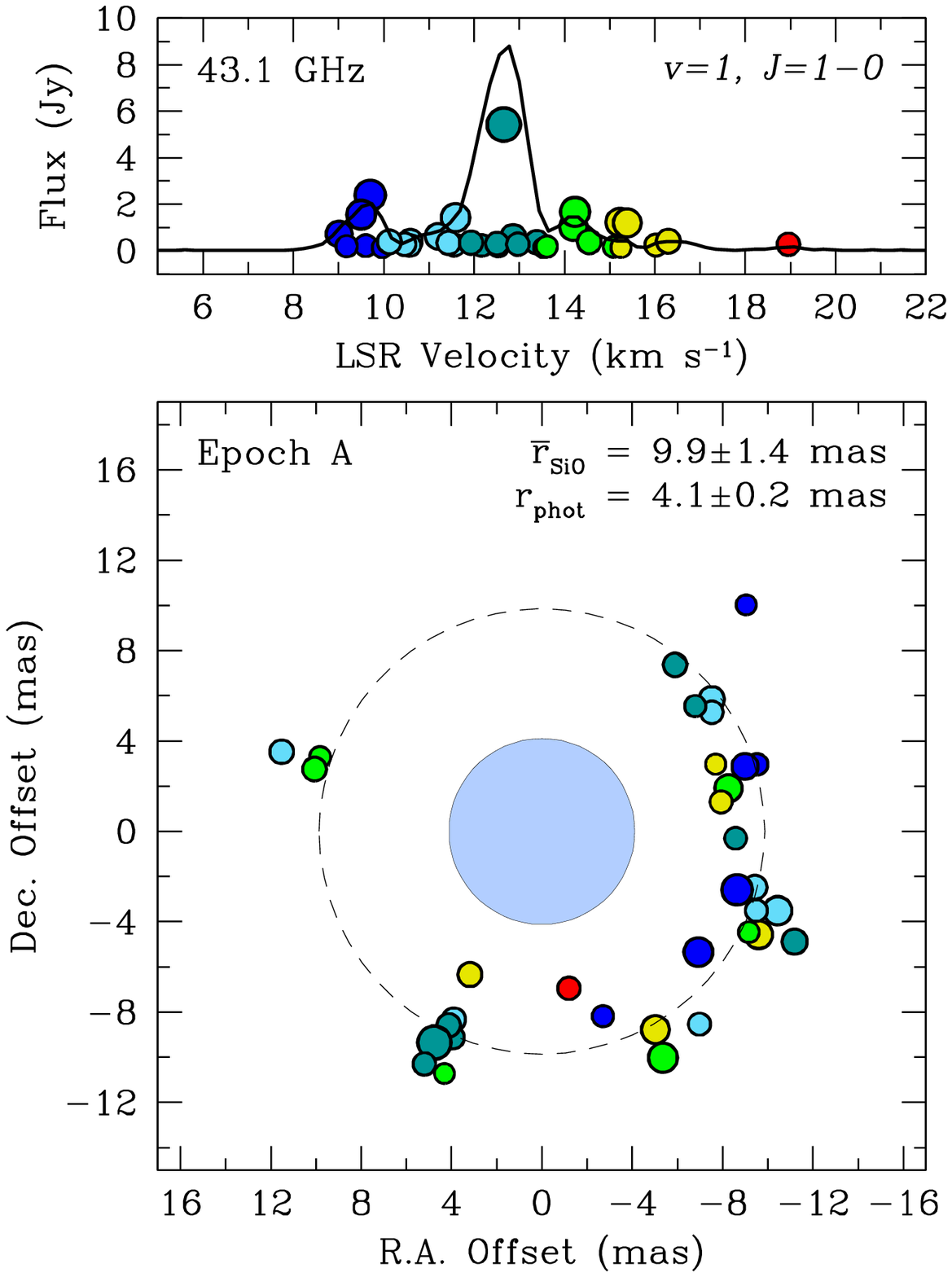}}
\resizebox{0.325\hsize}{!}{\includegraphics{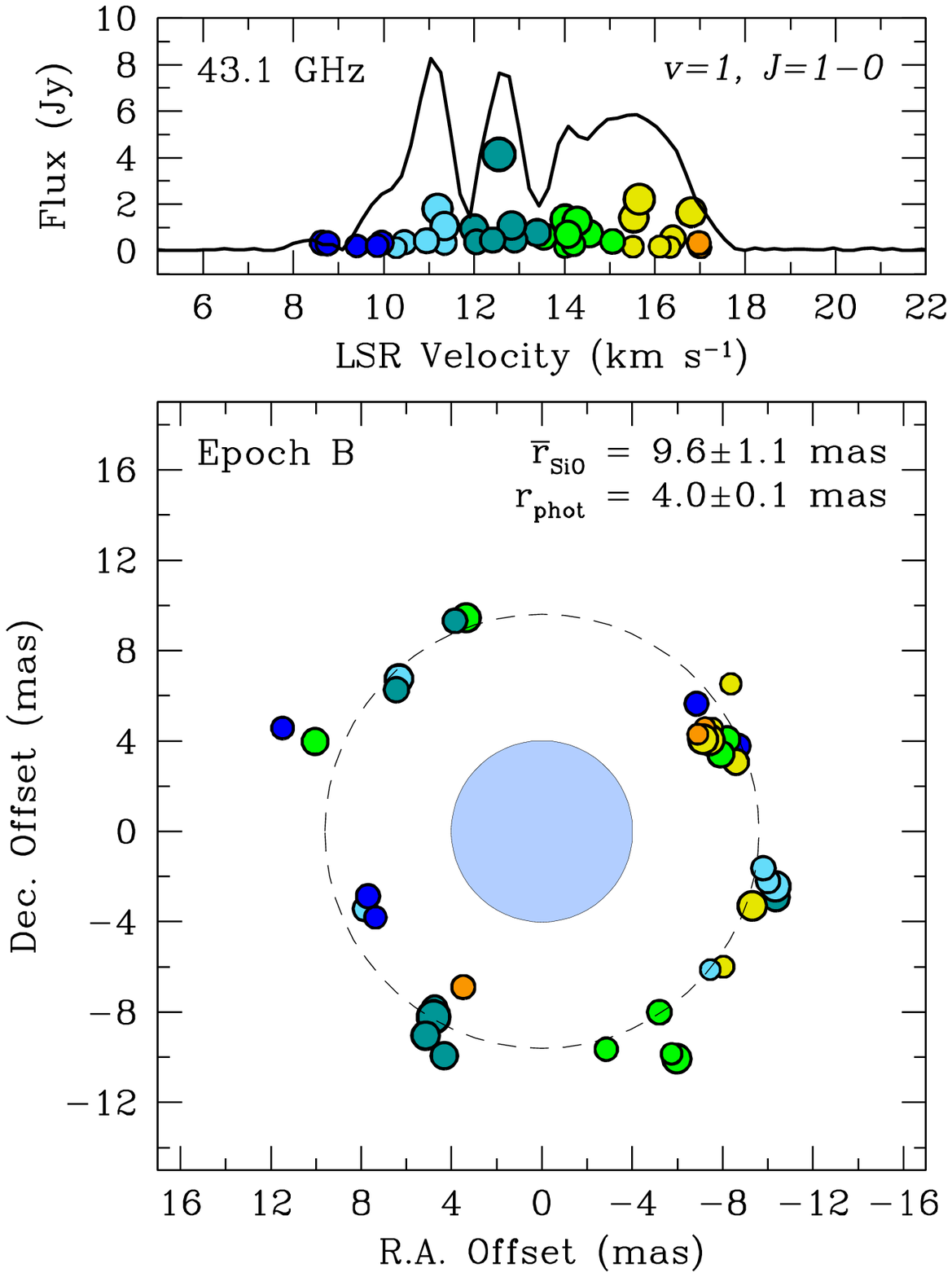}}
\resizebox{0.325\hsize}{!}{\includegraphics{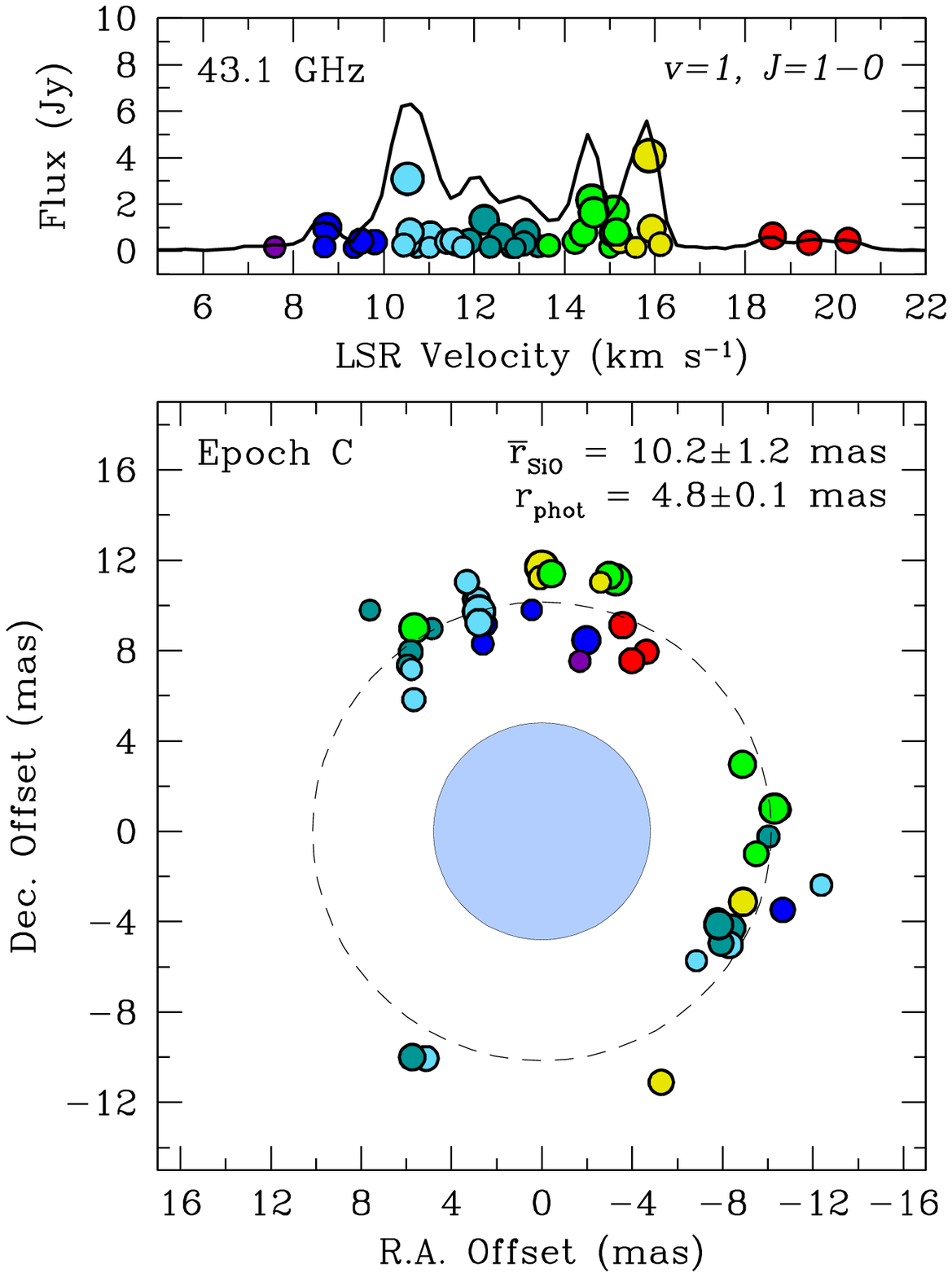}}

\resizebox{0.325\hsize}{!}{\includegraphics{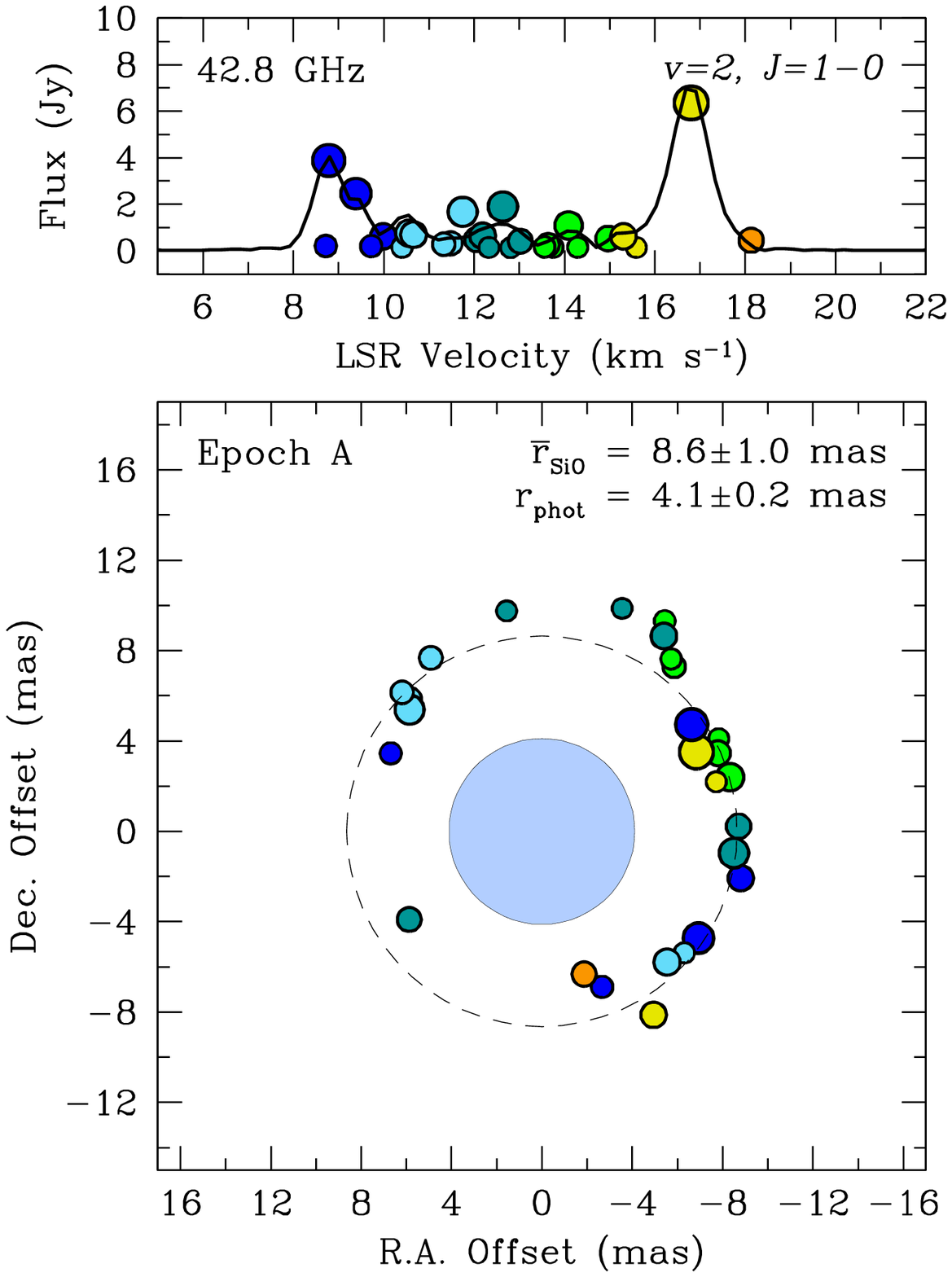}}
\resizebox{0.325\hsize}{!}{\includegraphics{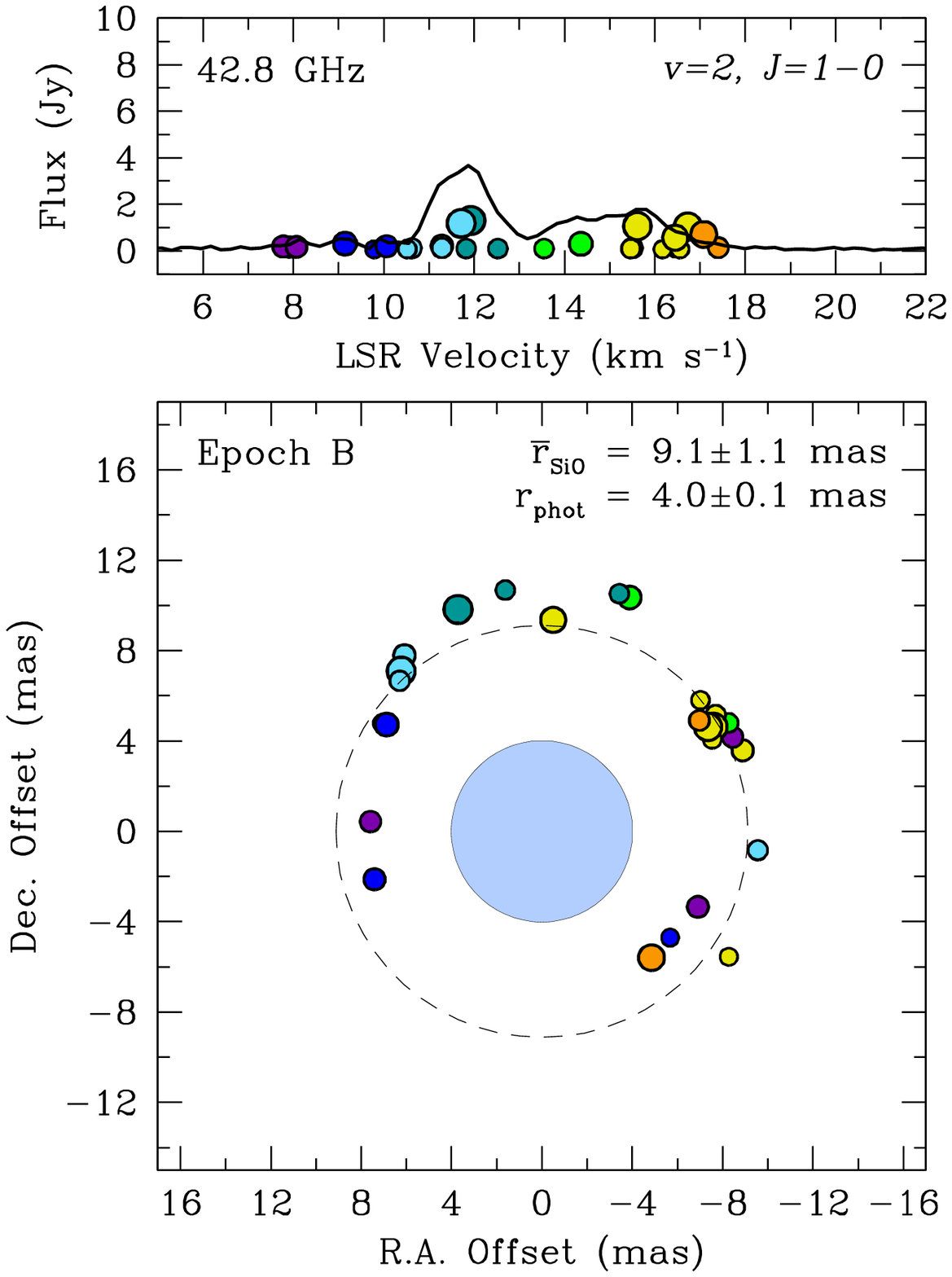}}
\resizebox{0.325\hsize}{!}{\includegraphics{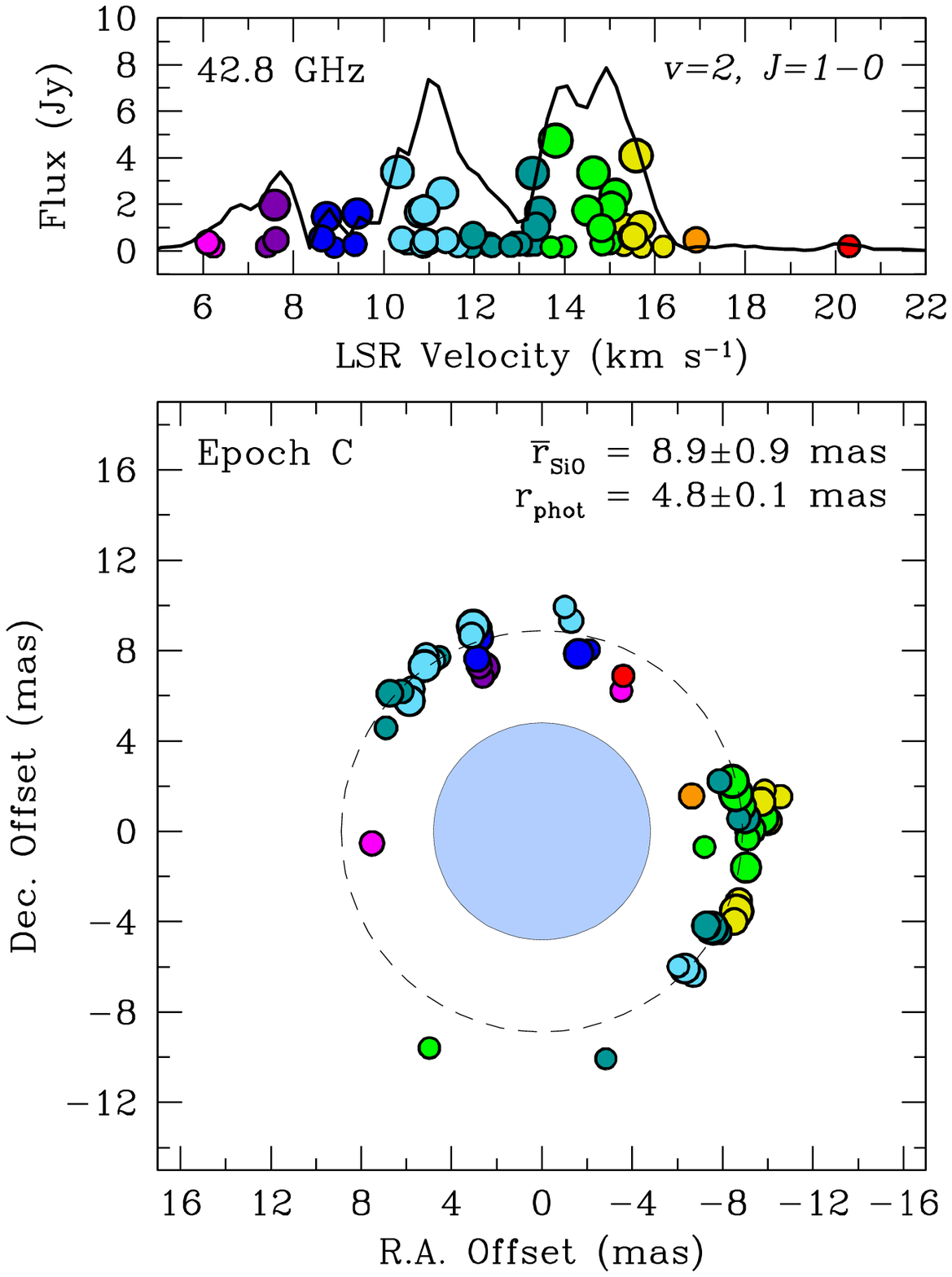}}
\caption{LOS velocity structure of the (top) $v=1, J=1-0$ (43.1\,GHz)
and (bottom) $v=2, J=1-0$ (42.8\,GHz) SiO maser emission
toward S~Ori. The epochs of observations are from left to
right Epoch A ($\Phi_\mathrm{vis}=0.46$), Epoch B ($\Phi_\mathrm{vis}=0.56$),
and Epoch C ($\Phi_\mathrm{vis}=1.14$). 
The top sub-panels of each epoch and transition show the spectra formed
by plotting maser intensity versus velocity, color-coded in
1.7~km~s$^{-1}$ velocity increments from redward to blueward.  
The solid lines in the top sub-panels represent the scalar-averaged
cross-power spectra averaged over all of the VLBA antennas.
The bottom (main) sub-panels plot the spatial and velocity
distributions of the masers. The color of each point represents
the corresponding velocity bin in the spectrum, and the size of each
point is proportional to the logarithm of the flux density.
Errors in the positions of the features are smaller than the data
points. The dashed circles are based on the mean angular distances of
the SiO masers from the centers of the distributions. 
The light blue filled circles 
in the centers illustrate for comparison the angular sizes of the 
continuum photosphere
derived in Sect.~\ref{sec:midimodelresults}. The true location of the 
star relative to the masers is unknown and here assumed to coincide with
the center of the maser spot distribution. We use also panels (e) of 
Figs.~\protect\ref{fig:midiA}--\protect\ref{fig:midiD}, as well 
as Fig.~\ref{fig:allepochs}, for a comparison
of the SiO maser ring radii to the extension of the molecular atmosphere
and the inner dust shell boundary.}
\label{fig:vlbavel}
\end{figure*}

The maser features characterized using the procedures described
above are listed in Tables \ref{tab:epochA_comps}--\ref 
{tab:epochC_comps} and are shown in Fig.~\ref{fig:vlbavel}, where   
there are six panels ordered by epoch left to right and by transition 
top to bottom.  Within each panel there are two 
sub-panels. Each upper sub-panel shows the spectrum of the maser emission 
with flux density plotted as a function of local standard of rest (LSR) 
velocity. Point sizes are proportional to the logarithm of the flux 
density and are color-coded according to velocity bin in increments 
of 1.7\,km\,s$^{-1}$. The bottom sub-panels of Fig.~\ref{fig:vlbavel} 
show the spatial distribution of the SiO masers as observed in the images. 
Point sizes are again proportional to the logarithm of the flux density 
with the same velocity color-coding as in the top sub-panels.

Examining the upper sub-panels and the component tables, we find that
the masers have shifted slightly toward the blue end of the spectrum
from our previous BW05 observations.  There the maser velocities
ranged from 11.1 to 24.7\,km\,s$^{-1}$, while for these latest three
epochs the masers span a velocity range from 8.6 to 19.0\,km\,s$^{-1}$.
The velocity range for our latest three epochs is consistent with  
earlier spectral measurements by Jewell et al. (\cite{jewell91}), who 
measured a velocity range of 7.1 to 19.6\,km\,s$^{-1}$ for the $v=1$ masers.
Spectral ranges measured by Cho et al. (\cite{cho96}) were much narrower
at 11.0--16.0\,km\,s$^{-1}$ for the $v=1$ and 12.0-14.7\,km\,s$^{-1}$
for the $v=2$ masers, respectively. For the three epochs discussed
here, flux densities of the peak component in the spectrum ranged
from 3.0 to 5.4\,Jy\,beam$^{-1}$ for the $v=1$ masers and 1.3 to
6.4\,Jy\,beam$^{-1}$ for the $v=2$ masers. These are close to our
BW05 results for which we measured peaks 
of 4.3~Jy~beam$^{-1}$ and 5.3~Jy~beam$^{-1}$ for the $v=1$ and $v=2$ masers, 
respectively.

When compared to the spectrally-averaged total-intensity images 
(Fig.~\ref{fig:allepochs}), the maser component maps in lower sub-panels of 
Fig.~\ref{fig:vlbavel} provide an accurate representation of the 
spatial structure of the maser emission summed over all velocity channels 
in the image cube. The component maps all show that the SiO masers in both 
transitions form partial to full rings of emission with the typical 
clumpy distribution. The top three panels show the $v=1, J=1-0$ to 
form a partial ring for epoch A with a gap to the north, nearly a full ring 
for epoch B , and again a partial ring dominated by emission in the 
northern and western regions for epoch C.  For the $v=2, J=1-0$ SiO masers 
shown in the bottom three panels, we observe nearly a full ring of emission
for epoch A, a partial ring with a gap to the south for epoch B,
and a partial ring dominated by northern and western features for
epoch C.  In all three epochs the $v=1$ masers appear to be distributed
over a larger area on the sky than the $v=2$ masers. In addition, the
$v=1$ rings appear thicker than the corresponding $v=2$ rings at each
epoch. 
It should be noted that, while epochs A and B were spaced only 42
days apart, there are significant differences in the number and
distribution of the masers in both transitions.  This is not unexpected
since previous multi-epoch VLBI studies
(e.g. Diamond \& Kemball \cite{diamond03}) have
shown the fine structure of the SiO masers to vary significantly over
periods as short as two weeks.  This variation, combined
with the relative complexity of the maser structures, precluded
us from tracking individual maser components in order to determine
maser proper motions.

To better characterize the apparent size and thickness of the SiO maser
distributions, we computed the average angular distance of the masers
from the center of their distribution ($\bar r$). To accomplish this, 
we first  determined the center of the distribution by performing a 
least-square fit of a circle to the combined $v=1$ and $v=2$ maser 
positions. We were able to combine the two sets of masers only because we had 
previously aligned the images of the two transitions following the procedure
described in Sect.~\ref{sec:vlba_reduction}. This fit produced a common center 
from which we computed the mean maser angular 
distance $\overline{r}_{\rm SiO}$ and the standard deviation
for each transition independently for each epoch.

The mean angular distances from the center for the observed SiO masers at
$v=1$, 43.1\,GHz and $v=2$, 42.8\,GHz are listed in 
Table~\ref{tab:maser_params}. These distances are also indicated by 
a dashed circle in each of the six panels in Fig.~\ref{fig:vlbavel}.  
The standard deviations in the mean distances are also listed in 
Table~\ref{tab:maser_params} and represented in Fig.~\ref{fig:vlbavel} 
as the error on the mean. Two times the 
standard deviation also provides an indication of the thickness of the shell.
The listed parameters in Table~\ref{tab:maser_params} are for each epoch 
of observation/visual phase and each of the  two transitions, 
the apparent mean angular distances from the center of the masers 
spots $\bar{r}_{v=1,2}$, the ratio of the 
mean distances of the two transitions $\bar{r}_{v=1}$/$\bar{r}_{v=2}$, 
as well as the mean LOS velocities $\bar{V}_{v=1,2}$ and their ratio 
$\bar{V}_{v=1}$/$\bar{V}_{v=2}$.
 
\begin{table}
\caption{Results from our VLBA observations of the 
$v=1$, 43.1\, GHz and $v=2$, 42.8\,GHz SiO maser transition toward S~Ori.}
\label{tab:maser_params}
\begin{tabular}{lrrr}
\hline\hline
Epoch                                 &  A  &  B  &  C   \\
Phase $\Phi_\mathrm{vis}$ & 0.46           & 0.56    & 1.14    \\\hline

$\bar{r}_{v=1}$ [mas] & 9.9 $\pm$ 1.4 & 9.6 $\pm$ 1.1 & 10.2 $\pm$  
1.2     \\
$\bar{r}_{v=2}$ [mas] & 8.6 $\pm$ 1.0 & 9.1 $\pm$ 1.1 & 8.9 $\pm$  
0.9     \\[1ex]

$\bar{r}_{v=1}$/$\bar{r}_{v=2}$  &  1.15  & 1.05  &  1.15     \\[1ex]

$\bar{V}_{v=1}$ [km\,s$^{-1}$] & 12.6 $\pm$ 2.4 & 13.3 $\pm$ 2.5 &  
13.0 $\pm$ 2.8    \\
$\bar{V}_{v=2}$ [km\,s$^{-1}$] & 12.5 $\pm$ 2.4 & 12.9 $\pm$ 3.2 &  
12.5 $\pm$ 2.9    \\[1ex]

$\bar{V}_{v=1}$/$\bar{V}_{v=2}$  &  1.01  & 1.03  &  1.04     \\
\hline
\end{tabular}
\end{table}

 From Table~\ref{tab:maser_params} we see that $\bar r$ ranges from
8.6 to 10.2\,mas. At every epoch we find that the $v=1$ masers have a
greater mean distance from center than the $v=2$ masers. This finding 
is consistent with BW05 in which we found $\bar r_{v=1} = 9.4$\,mas 
and $\bar r_{v=2} = 8.8$\,mas. In BW05, however, we were unable to make a
definitive statement about the relative sizes of the two maser distributions 
because we had only the one epoch of observations, and at this epoch, 
the $v=2$ masers were primarily confined to a small region on the northwest 
side of the shell. In our more recent epochs, the rings of emission are much 
more complete making the comparison between transitions easier. Although the  
differences in $\bar r$ for epochs A, B, and C are less than the standard 
deviation (thickness of the ring), they are still greater than the resolution
of the images given earlier by the synthesized beam sizes 
of $0.6 \times 0.4$~mas for epoch A and $0.5\times 0.2$~mas for 
epochs B and C.  
The fact that the $v=1$ ring is consistently larger than the $v=2$ ring 
suggests that  for S~Ori the two sets of masers are not co-spatial.

The relative separation of the $v=1$ and $v=2$ masers for S~Ori is also 
consistent with recent results obtained for the SiO masers around other stars  
including: \object{IRC~ +10011} (Desmurs et al. \cite{desmurs00});  
\object{TX~Cam} (Desmurs et al. \cite {desmurs00}, 
Yi et al. \cite{yi05}); $o$~Cet, \object{U~Ori}, and \object{R~Aqr} 
(Cotton et al. \cite{cotton06}).
In addition, numerical simulations of SiO masers
(Humphreys et al.~\cite{humphreys96}; Gray \& Humphreys~\cite{gray00})
show the $v=1, J=1-0$ masers to occur farther from the star than the
$v=2, J=1-0$ masers throughout the entire pulsation. The simulations
of Gray \& Humphreys~(\cite{gray00}) also show that the mean thickness
of the $v=1$ shell is roughly twice that of the $v=2$ shell throughout the 
stellar cycle. In our measurements, we find the $v=2$ shell to be thicker  
in the BW05 observations, the $v=1$ shell to be thicker in epochs A and  
C, and the two transitions to be roughly equivalent in epoch B. In the two
cases for which the $v=1$ shell was thicker, the ratio of the two thicknesses
is 1.3--1.4, much less than the 2 determined by Gray \& Humphreys.
The angular sizes of the maser shells are reported in 
Table~\ref{tab:results}.  At the  assumed distance
for S~Ori of $480\ \mathrm{pc}\ \pm\ 120\ \mathrm{pc}$ (van Belle et  
al. \cite{vanbelle02}), 1\,mas is roughly equivalent to 0.48~AU. 
Thus the linear-scale sizes  for maser distributions range from 8.3 to 9.8\,AU 
and the maser shell thickness (twice the standard deviation) ranges from 
roughly 0.9 to 1.3 AU.
\subsubsection{The kinematics of the SiO masers}
In BW05, we reported that the SiO masers showed no coherent  
velocity structure indicative of global expansion/infall or rotation.  
Examining Fig.~\ref{fig:vlbavel}, we again find no evidence of such 
coherence in the line-of-sight (LOS) velocities of the masers as a 
function of spatial location.  We did, however, notice that there
appears to be a velocity gradient at all epochs, with masers toward  
the blue- and red-shifted ends of the spectrum lying closer to the center 
of the  distribution than masers at intermediate velocities. This phenomenon 
was also found in BW05 and has been previously observed in other SiO maser 
studies (e.g. Boboltz \& Marvel \cite{boboltz00}; 
Hollis et al. \cite {hollis01}).

\begin{figure}
\centering
\resizebox{1.\hsize}{!}{\includegraphics{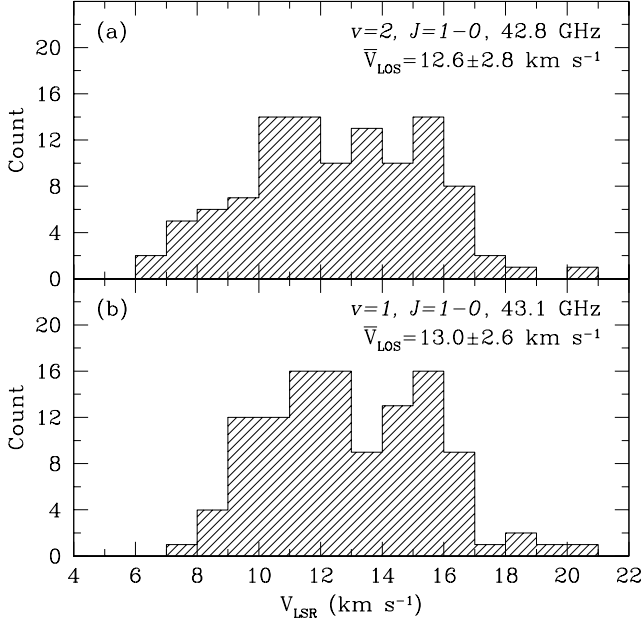}}
\caption{Histogram of the LOS velocities of our SiO maser spots combining
VLBA epochs A-C. Panel (a) shows the histogram for the $v=2$ 42.8\,GHz
masers and panel (b) for the $v=1$ 43.1\,GHz masers. Also indicated
are the mean values $\bar V_{v=1}$ and  $\bar V_{v=2}$. 
$\bar V_{v=1}$ is slightly redder than $\bar V_{v=2}$.}
\label{fig:velhistl}
\end{figure}

The LSR stellar velocity of S~Ori has not been measured particularly well. 
In the
General Catalog of Stellar Radial Velocities (Wilson \cite{wilson53}),
the value listed for S~Ori is 22\,km\,s$^{-1}$. Young (\cite{young95}) 
observed the sub-millimeter CO (3-2) and CO (4-3) lines
toward S~Ori and determined values of $V_{\rm LSR} = 14.5 \pm 0.2$ and
$14.1 \pm 0.5$\,km\,s$^{-1}$ from fits to the spectra, respectively.
Winters et al. (\cite{winters03}) derived $V_{\rm LSR} = 14.0$\,km\,s$^{-1}$
based on fits of the CO (1-0) and CO (2-1) lines. 
Since the SiO masers are not distributed evenly about any of the above
velocities, we computed an average LOS velocity from the masers
themselves for each transition at epoch.
The values of $\bar V_{\rm LOS}$ for each of the epochs are reported 
in Table~\ref{tab:maser_params} with errors of 15\% estimated from the 
uncertainty of the absolute amplitude calibration.   
Although the differences  are well within the error bars, it appears that 
the $v=2$ masers are shifted slightly toward the blue end of the spectrum 
relative to the $v=1$ masers at all three epochs. In addition
to computing values for individual transitions and epochs, we also  
combined the three epochs and computed a mean LOS velocity for each 
transition. 
We obtained mean values of 
$V_{\rm LSR} = 12.6 \pm 2.8$\,km\,s$^{-1}$ and 
$13.0 \pm 2.6$\,km\,s$^{-1}$ for the
42.8\,GHz and 43.1\,GHZ masers, respectively.
These values are shown in 
Fig.~\ref{fig:velhistl}.  Again we see that $\bar V_{v=1}$ is slightly 
redder than $\bar V_{v=2}$.

\begin{figure}
\centering
\resizebox{1.\hsize}{!}{\includegraphics{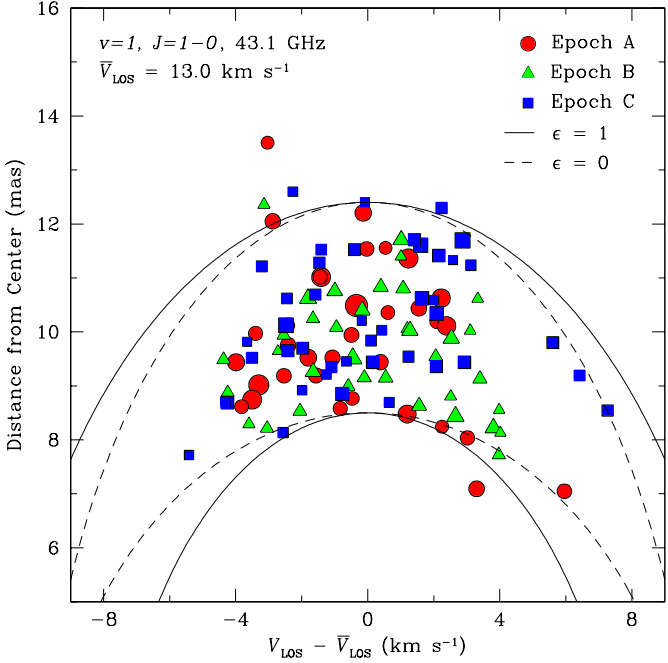}}
\caption{Distances of the 43.1\,GHz maser components from the common
center of the distribution (radius) versus their velocity relative to the 
average LOS velocity for the 43.1\,GHz 
transition ($V_{\rm LOS} - \bar V_{\rm LOS}$).
Maser components of the different epochs are denoted by different symbols
and colors. The plot indicates that higher-velocity masers are found closer 
to the star. This trend can be explained by means of expanding maser shells.
The curves represent shells at the inner and outer boundaries
of the SiO maser shell using (solid lines) a logarithmic velocity gradient
and (dashed lines) constant velocity expansion. See text for details.
}
\label{fig:43.1_radvsvel}
\end{figure}

\begin{figure}
\centering
\resizebox{1.\hsize}{!}{\includegraphics{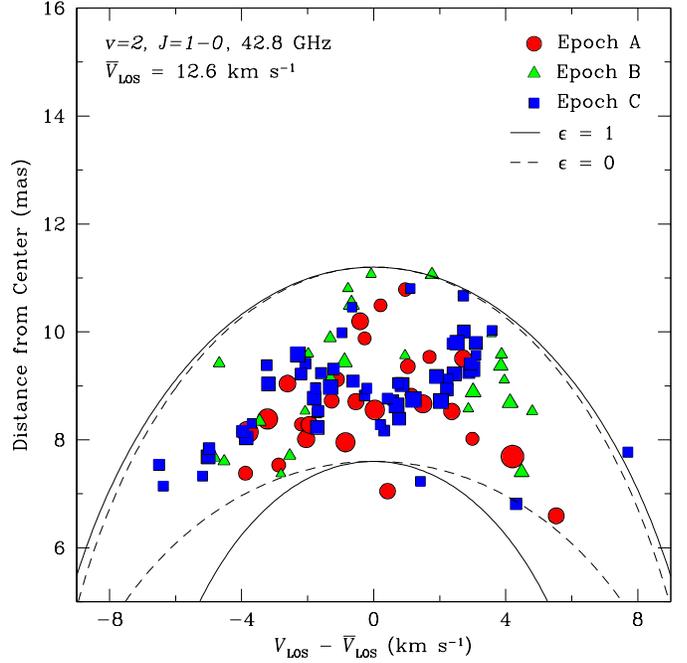}}
\caption{As Fig.~\protect\ref{fig:43.1_radvsvel}, but for the 42.8\,GHz 
transition. The axes' scales are identical for the two figures. 
Compared to Fig.~\protect\ref{fig:43.1_radvsvel}, it is evident that
the 42.8\,GHz maser spots lie at systematically closer distances to the
center of the common maser distribution than the 43.1\,GHz maser spots.} 
\label{fig:42.8_radvsvel}
\end{figure}

To verify the notion that the higher-velocity masers are found closer  
to the star, we plotted component distance from the center of the 
distribution  (radius) versus its velocity relative to the average 
LOS velocity for that transition ($V_{\rm LOS} - \bar V_{\rm LOS}$).  
The results are shown in Figs.~\ref{fig:43.1_radvsvel} 
and \ref{fig:42.8_radvsvel} for the  $v=1, J=1-0$ 43.1\,GHz and 
the $v=2, J=1-0$ 42.8\,GHz transitions, respectively. In the figures we  
plotted each epoch with a different symbol.  The masers in both transitions  
appear to have the same distribution with a central peak near
$V_{\rm LOS} - \bar V_{\rm LOS} = 0$ and decreasing maser radius as
a function of increasing values of $|V_{\rm LOS} - \bar V_{\rm LOS}|$.

A simple model that is often used in the case of OH (e.g. Reid et  
al. \cite{reid77}) and H$_2$O (e.g. Yates \& Cohen \cite{yates94}) maser
kinematics is that of a uniformly expanding thin shell.  In this  
model the projected distance from center ($a$) of a maser on the shell is  
related to its LOS velocity $V - V_{\star}$ by the expression:
\begin{equation}
\biggl(\frac{a}{r_{\rm sh}}\biggr)^2 +
\biggl(\frac{V - V_{\star}}{V_{\rm sh}}\biggr)^2 = 1,
\end{equation}
where $r_{\rm sh}$ and $V_{\rm sh}$ are the radius and the
expansion velocity of the shell, respectively. The LSR velocity of the
star ($V_{\star}$) is assumed here to be equal to $\bar V_{\rm LOS}$.
Numerical simulations of SiO masers often employ the Sobolev or large 
velocity gradient (LVG) approximation (Lockett \& Elitzur \cite{lockett92}; 
Bujarrabal \cite{bujarrabal94}; Doel et al. \cite{doel95}; 
Humphreys et al \cite{humphreys02}). The LVG approximation assumes a 
large velocity gradient across the SiO maser region, thus producing large 
Doppler shifts that serve to disconnect a region of velocity coherence from 
the surrounding medium (Bujarrabal \cite {bujarrabal94}).
The parameter typically used to characterize the velocity field in the
maser region is the logarithmic velocity gradient $\epsilon$ given by
\begin{equation}
\epsilon =  \frac{d \ln V}{d \ln r} = \frac{r}{V} \frac{dV}{dr}.
\end{equation}
A value of $\epsilon = 0$ is equivalent to constant velocity expansion,
while $\epsilon \ge 1$ corresponds to a velocity field with large  
radial accelerations. The SiO maser numerical simulations 
(e.g. Lockett \& Elitzur \cite {lockett92}, Bujarrabal \cite{bujarrabal94}, 
Doel et al. \cite{doel95}) typically  use a value of $\epsilon = 1$. 
Chapman \& Cohen (\cite{chapman86}) determined values of $\epsilon$ for 
the various maser species surrounding the supergiant \object{VX~Sgr}
and found $\epsilon \approx 1$ in the SiO region, $\epsilon \approx  
0.5$ in the region of the H$_2$O and mainline (1665\,MHz, 1667\,MHz)
OH masers, and $\epsilon \approx 0.2$ in the 1612 MHz OH maser region.

Plotted along with the data in Figs.~\ref{fig:43.1_radvsvel} and
\ref{fig:42.8_radvsvel} are four curves computed using the above  
equations. They represent shells at the inner and outer 
boundaries of the SiO masers.
For the masers in each transition, we used an inner 
shell radius equivalent to the smallest average distance minus its 
standard deviation from Table~\ref{tab:maser_params}.  
Similarly, for the outer shell radius we used the largest average distance 
plus its standard deviation from Table~\ref{tab:maser_params}.   
For the scenario of a logarithmic velocity gradient, 
the velocity at the inner $v=2$ boundary was set at 7\,km\,s$^{-1}$ and 
the velocities of the other shells computed assuming $\epsilon = 1$.  
The velocity of the outermost shell of the $v=1$ transition 
is $V_{\rm sh} \approx 10.5$\,km\,s$^{-1}$.

For the scenarion of a constant velocity expansion ($\epsilon = 0$),
the best by-eye fit results in $V_{\rm sh} = 10$\,km\,s$^{-1}$.  
In this case, the higher expansion velocity at the inner boundary of 
the $v=2$ transition is
required in order to provide a reasonable fit to the outer boundaries of
both transitions.  For the $\epsilon = 1$ case, the inner $v=2$ velocity
can be lower with the radial acceleration allowing for a better
fit to the outer boundaries.  For comparison, the escape velocities  
computed for the innermost  $v=2$ and outermost $v=1$ shells, assuming
a mass for S~Ori of 1.0~$M_{\odot}$, are 22.1 and 18.0\,km\,s$^{-1}$,  
respectively. Therefore, even at the velocities bracketed by the curves, 
the maser gas is still gravitationally bound.

From the SiO maser data on S~Ori, we cannot definitively rule out constant
velocity expansion; however, the SiO maser region is very dynamic,  influenced
by the pulsation of the star and the passage of shocks.  Theoretical studies 
all seem to include some form of velocity gradient across the SiO maser  
region. In addition, constant velocity expansion would be more conducive to  
masers along the line of sight to the star because of the increased velocity  
coherence. SiO masers are rarely observed in front of the stellar disc, and
for S~Ori in particular, such masers have not be seen.  We therefore  
consider some form of acceleration in the region the more likely scenario.

The canonical interpretation of the rings formed by the SiO masers is  
that the masers are tangentially amplified and primarily confined to a region
close to the plane of the sky.  Under this interpretation, the  
ring size and thickness are direct indicators of the equivalent 3-D shell size 
and thickness.  The radially expanding spherical shell described above 
will naturally result in a projection effect where masers out of the plane 
of the sky have smaller angular distances from center due to the projection.  
In the case of S~Ori, this projection is rather small. The equation for the 
angle $\theta$ measured between the LOS and a point on the shell locating 
the maser can be written in terms of the velocities as
\begin{equation}
\sin \theta = \biggl[1 - \biggl( \frac{V}{V_{\rm sh}} \biggr)^2  
\biggr]^{1/2}.
\end{equation}
In the most extreme case, all of the thickness observed in the ring is
due to projection of the masers on the  outermost thin shell with a  
velocity of $V_{\rm sh} \approx 10.5$\,km\,s$^{-1}$ ($\epsilon = 1$).
Figures~\ref{fig:velhistl}--\ref{fig:42.8_radvsvel} show that the  
majority of the masers are within $\sim$\,4\,km\,s$^{-1}$ of LOS 
center velocity.
For these extreme 4\,km\,s$^{-1}$ masers, $\sin \theta \approx 0.92$ and
$\theta \approx 67^{\circ}$.  This results in a difference of $\sim$\,8\%
between the projected distance of the maser and its true radial  
distance. For masers less than 4\,km\,s$^{-1}$ the difference is smaller.   
Assuming the masers are distributed uniformly, the differences between
the projected ring sizes and thicknesses and their 3-D equivalents is
$\sim$4\%.  Thus, the given errors of the projected ring sizes determined 
from the maser distributions include the uncertainty of the projection effect.
\section{VLTI/MIDI and VLBA results as a function of stellar phase}
\label{sec:comparison}
\begin{table}
\caption{Overview of the photometric and
spatial parameters of S~Ori as a function of stellar phase,
derived from both our VLTI and VLBA observations.
}
\label{tab:results}
\begin{tabular}{lrrrr}
\hline\hline
                               &  A  &  B  &  C  &  D  \\\hline
Phase $\Phi_\mathrm{vis}$ & 0.44           & 0.55    & 1.15     & 1.27  \\[1ex]
$V$ [mag]         & 12.8 $\pm$ 0.3 & 14.0 $\pm$ 0.3 & 9.4 $\pm$ 0.3 & 10.5 $\pm$ 0.4 \\
$f_N$ [Jy]        & 187 $\pm$ 12   & 152 $\pm$ 17 & 198 $\pm$ 18 & 203 $\pm$ 13 \\
$f_{43.1}$ [Jy] & 119 $\pm$ 18  & 203 $\pm$ 30 & 198 $\pm$ 30  &   \\
$f_{42.8}$ [Jy] & 148 $\pm$ 22  &  46 $\pm$  7 & 383 $\pm$ 57  &   \\[1ex]

$\Theta_\mathrm{Phot}$ [mas] & 9.0 $\pm$ 0.3 & 7.9 $\pm$ 0.1 & 9.7 $\pm$ 0.1 & 9.5 $\pm$ 0.4  \\[1ex]

$\Theta_\mathrm{in}$ [mas] & 16.2 $\pm$ 1.9  & 15.9 $\pm$ 1.6 & 21.2 $\pm$ 1.9 & 22.8 $\pm$ 2.1  \\[1ex]
$\Theta_{43.1}$ [mas] & 19.6 $\pm$ 2.8 & 19.0 $\pm$ 2.6 & 20.4 $\pm$ 2.4  &     \\
$\Theta_{42.8}$ [mas] & 19.0 $\pm$ 2.0 & 17.8 $\pm$ 3.4 & 17.8 $\pm$ 1.8  &     \\[1ex]

$\Theta_\mathrm{in}/\Theta_\mathrm{Phot}$ & 1.8 $\pm$ 0.2 & 2.0 $\pm$ 0.2 & 2.2 $\pm$ 0.2  & 2.4 $\pm$ 0.2     \\[1ex]

$\Theta_{43.1}/\Theta_\mathrm{Phot}$ & 2.2 $\pm$ 0.3 & 2.4 $\pm$ 0.3 & 2.1 $\pm$ 0.3 &    \\
$\Theta_{42.8}/\Theta_\mathrm{Phot}$ & 2.1 $\pm$ 0.2 & 2.3 $\pm$ 0.4 & 1.9 $\pm$ 0.2 &    \\[1ex]

$\Theta_{43.1}/\Theta_\mathrm{in}$ & 1.2 $\pm$ 0.2 & 1.2 $\pm$ 0.2 & 1.0 $\pm$ 0.1 &  \\
$\Theta_{42.8}/\Theta_\mathrm{in}$ & 1.2 $\pm$ 0.2 & 1.1 $\pm$ 0.2 & 0.8 $\pm$ 0.1 &  \\[1ex]

$R_\mathrm{Phot}$ [R$_\odot$]  & 469 $\pm$ 127 & 411 $\pm$ 111 & 498 $\pm$ 134  &  491 $\pm$ 133 \\\hline
\end{tabular}
\end{table}
Table~\ref{tab:results} provides an overview on the 
photometric and spatial parameters of S~Ori as a function
of observational epoch/stellar phase, as derived from both
our VLTI/MIDI and VLBA observations in Sects.~\ref{sec:vlti}
and \ref{sec:vlba}. 
The listed photometric parameters include the $V$ magnitude
based on the lightcurve in Fig.~\protect\ref{fig:lightcurve}, 
the $N$-band flux  based on our photometric MIDI data 
(Sect.~\ref{sec:midireduction}), and the total flux values of 
the two maser transitions (Sect.~\ref{sec:vlba_reduction}).
The listed spatial information includes the photospheric
angular diameter $\Theta_\mathrm{Phot}$ 
(Sect.~\ref{sec:midimodelresults}), the inner dust shell
diameter $\Theta_\mathrm{in}$ (Sect.~\ref{sec:midimodelresults}), 
and the maser ring angular diameters $\Theta_{43.1}$ 
and $\Theta_{42.8}$ (Sect.~\ref{sec:vlba_spatial}). 
To compare the continuum photospheric diameter, the inner 
dust shell diameter, and the maser shell diameters, the ratios
$\Theta_\mathrm{in}/\Theta_\mathrm{Phot}$,
$\Theta_{43.1/42.8}/\Theta_\mathrm{Phot}$, and 
$\Theta_{43.1/42.8}/\Theta_\mathrm{in}$ are listed as well. 
As an absolute scale, the stellar photospheric 
radius $R_\mathrm{Phot}$ at each phase is derived
from $\Theta_\mathrm{Phot}$ and the adopted distance 
to S~Ori (Sect.~\ref{sec:lightcurve}).
Here, the variability
phases are the mean values of the respective phases of 
the MIDI and VLBA observations. 

The continuum photospheric radii $R_\mathrm{Phot}$, 
the inner dust shell radii $R_\mathrm{in}$ 
(corresponding to $\Theta_\mathrm{in})$ 
and the maser shell radii $R_{43.1}$ and $R_{42.8}$
(corresponding to $\Theta_{43.1/42.8}$) are indicated
in  panels (d) of Figs.~\ref{fig:midiA}-\ref{fig:midiD},
which show the synthetic mid-infrared CLVs. This
allows a comparison of these radii to each other and to the 
profile of the extended molecular atmosphere.
Figure~\ref{fig:allepochs} provides as well a 
graphical comparison of the photospheric disc, the molecular layers, 
the dust shell, and the maser spots for each epoch. 
\paragraph{Variability of visual, mid-infrared, and maser flux densities}
The visual, mid-infrared, and maser fluxes all show
clear variability as a function of observational epoch.
The stellar phases given throughout this paper are visual 
phases as introduced in Sect.~\ref{sec:lightcurve}. 
The $N$-band flux $f_N$ shows clear variability
with an amplitude of about 50\,Jy ($\sim$ 30\%). 
$f_N$ exhibits a minimum at our epoch B at visual 
phase 0.55, i.e. shortly beyond the visual minimum,
rises toward stellar maximum, and continues to increase 
between epochs C and D at visual phases 1.15 and 1.27, 
probably indicating a phase lag and different shape of 
the $N$-band variability curve compared to the visual lightcurve.
The total maser fluxes are significantly lower near stellar
minimum (epochs A and B) compared to stellar maximum (epoch C), with
the exception of the strong $f_{43.1}$ value at epoch B. The latter
can most likely be explained by the larger number of maser spots
forming a nearly complete ring-like structure for this transition 
and epoch, while the other transitions and epochs show only partial
maser rings. 
\paragraph{Photospheric and dust shell radii as a function of visual phase}
The photospheric angular diameter $\Theta_\mathrm{Phot}$ derived from
our MIDI observations and modeling shows a clearly phase-dependent
size with an amplitude of about 1.8\,mas ($\sim$\,20\%) that
is well-correlated in phase with the visual lightcurve. 
It shows a minimum at epoch B, which is nearest to the visual 
minimum, and a maximum at epoch C, which is nearest to the visual maximum. 
Also the inner dust-shell radius shows a variation 
(amplitude $\sim$\,7\,mas $\sim$\,30\%), which is more closely 
correlated in phase and amplitude to $f_N$ than to $V$. The inner 
dust shell radius
is smallest at our epoch B near visual minimum and largest at epoch D
beyond visual maximum (phase 1.27).
Compared to $\Theta_\mathrm{Phot}$, the inner dust shell boundary is
located significantly closer to the stellar surface at our epochs A and B near 
visual minimum (1.8--2.0\,$\times\Theta_\mathrm{Phot}$) than at the 
post-maximum visual phases C and D (2.2--2.4\,$\times\Theta_\mathrm{Phot}$). 
Simultaneously, as discussed in Sect.~\ref{sec:dust}, the density 
gradient $p$ is significantly steeper and the optical depth significantly 
higher near visual minimum than at the post-maximum visual phases. This
can be understood by a wind model with increased mass loss near minimum 
phase occurring close to the stellar surface and an expanded dust shell after
maximum (cf. Sect.~\ref{sec:midimodelresults}).
\paragraph{Maser ring radii as a function of visual phase}
The mean SiO maser ring radii obtained at our three VLBA epochs
show a small variability with amplitude of $\le 7\%$, which is 
only a fraction of the mean width of the maser shells of $\sim 13\%$ and 
clearly below the variability amplitude of the photospheric and dust 
shell radii at the same epochs. However, we cannot rule out that the 
maser ring radii show a greater variability in between our epochs
of observation. Observations by 
Cotton et al. (\cite{cotton06}) consistently find variations in 
maser ring diameters of 3--14\% for different Mira stars, 
but no clear correlation with visual variability phase.
Theoretical estimates by Humphreys et al. (\cite{humphreys02})
predict a variability of the 43.1\,GHz maser shell radius 
with an amplitude of $\sim$\,20\%.

\paragraph{Maser ring radii compared to the photospheric radius}
The true location of the star relative to the maser spots is unknown.
Based on the assumption that the center of the maser spot distribution
coincides with the center of the star, a comparison of the maser ring
radii with the continuum photospheric radii gives an estimate of the distance
of the maser spots from the stellar photosphere.
The ratios between the maser ring radii and the photospheric radii
in Table~\ref{tab:results} range between 2.1$\pm$0.3 and 2.4$\pm$0.3.
These values are consistent with the values reported in BW05 (1.9 and 2.0),
with those by Cotton et al. (\cite{cotton04}), which range for 
different Mira stars --not always compared at the same phase-- between 
1.8 and 2.9, and the value by Fedele et al. (\cite{fedele05}) for R~Leo 
of 1.8. 
Our observations indicate a variability in the ratio 
of the mean maser shell radius to the photospheric radius 
with an amplitude of $\sim$\,10\% for the 43.1\,GHz transition, and 
$\sim$\,20\% for the 42.8\,GHz transition, with uncertainties on the
same order. This observed variability is dominated by the 
variability of the stellar continuum photospheric diameter.
\paragraph{Maser ring radii compared to the molecular atmosphere}
Panels (d) of Figs.~\ref{fig:midiA}--\ref{fig:midiD} as well 
as Fig.~\ref{fig:allepochs} show that the
mean maser ring radii at all our epochs mark the region just beyond 
the steepest decrease in the mid-infrared model intensity, 
i.e. shortly outward of the densest region of the molecular atmosphere.
\paragraph{Maser ring radii compared to the inner dust shell boundary} 
Table~\ref{tab:results}, as well as panels (d) of 
Figs.~\ref{fig:midiA}--\ref{fig:midiD}, indicate that the inner boundary
of the Al$_2$O$_3$ dust grains near visual minimum (epochs A and B)
appears to be located between the steepest decrease in the 
mid-infrared model intensity and the mean location of the SiO maser spots.
This means that the low-density atmospheric molecular layers, 
the SiO maser spots, and the Al$_2$O$_3$ dust grains are co-located near
visual minimum. At our post-maximum epochs C and D, the inner boundary of the 
Al$_2$O$_3$ dust shell appears significantly expanded, while the 
maser shells remain at about the same location. As a result, the 
inner boundary of the Al$_2$O$_3$ dust grains appears to have expanded
outward of the mean SiO maser rings at the post-maximum phase.
\section{Summary, conclusions, and discussion}
\begin{figure*}
\centering
\hfill%
\resizebox{0.48\hsize}{!}{\includegraphics{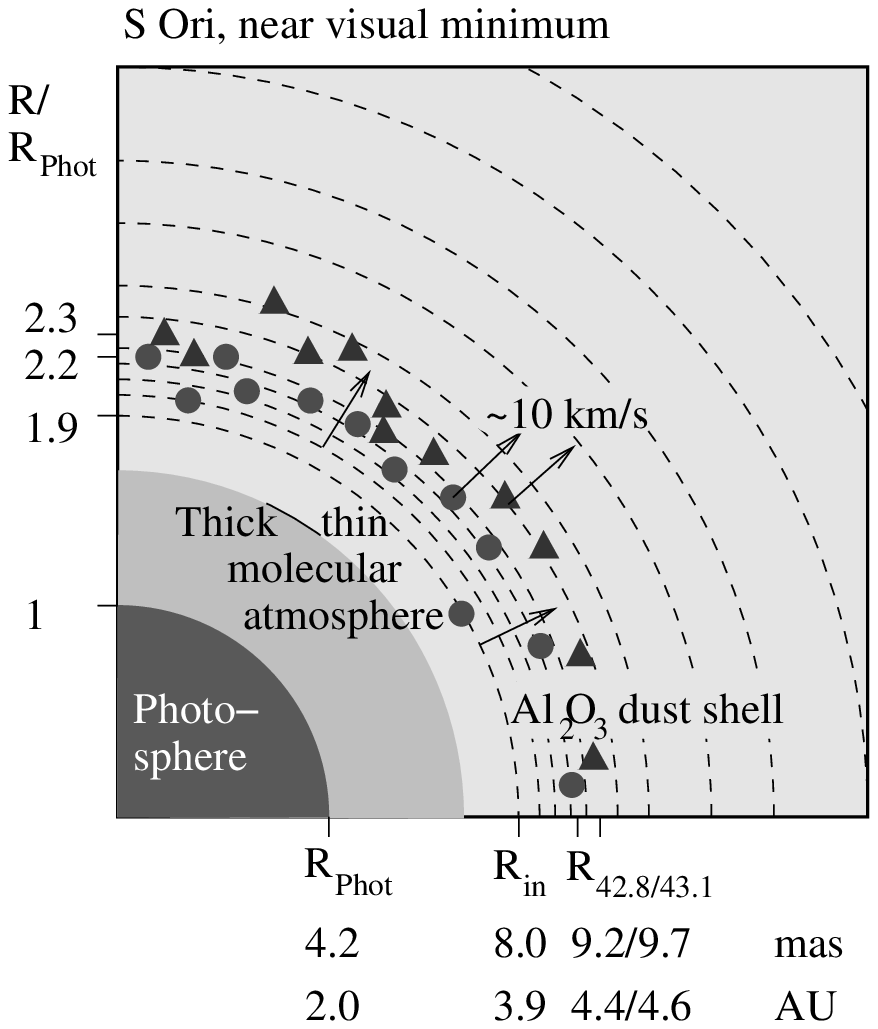}}
\hfill%
\resizebox{0.48\hsize}{!}{\includegraphics{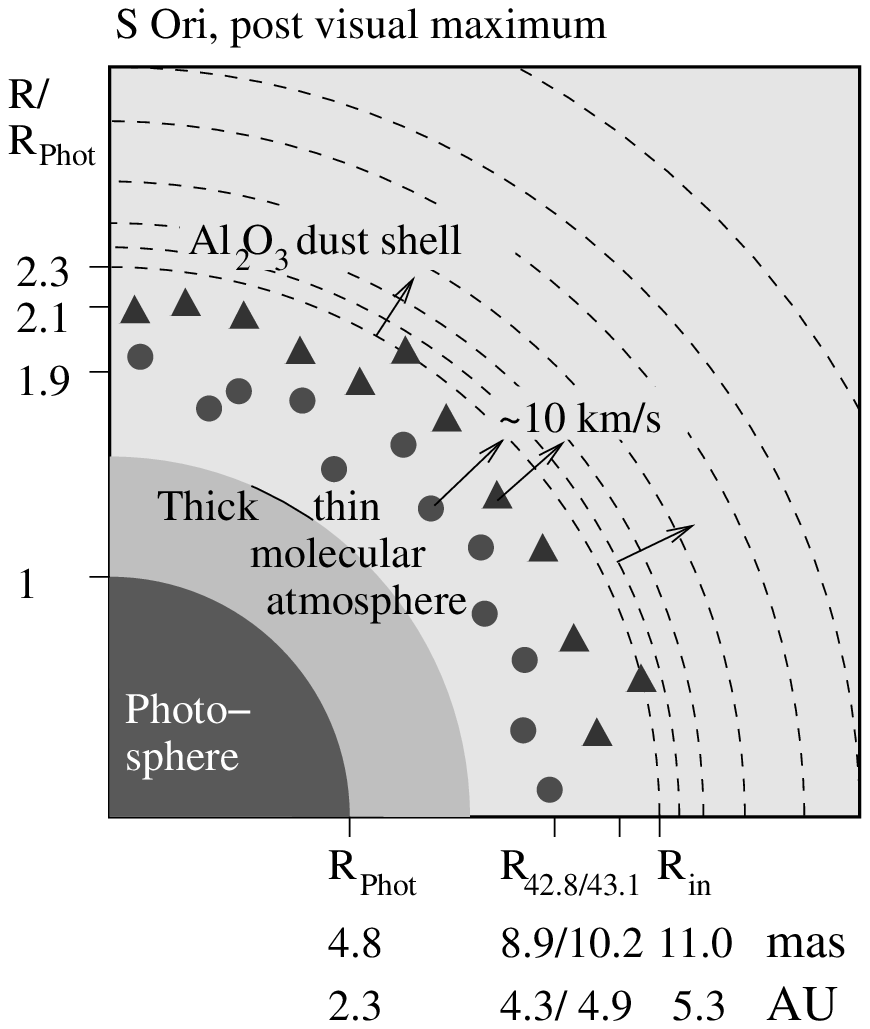}}
\hfill%
\caption{Sketch of the radial structure of S~Ori's CSE at (left) 
near-minimum and (right) post-maximum visual phase as derived in this work. 
Shown are the locations of the continuum photosphere (dark gray), 
the at $N$-band optically thick molecular atmosphere (medium dark gray), 
the at $N$-band optically thin molecular atmosphere (light gray), 
 the Al$_2$O$_3$ dust shell (dashed arcs), and  
the 42.8\,GHz and 43.1\,GHz maser spots (circles/triangles). The numbers 
below and beside the panels are the mean values of (left) 
epochs A \& B and (right) epochs
C \& D from Table~\protect\ref{tab:results}.}
\label{fig:sketch}
\end{figure*}
We have obtained the first time series of observations of a Mira 
variable star that include concurrent infrared and radio interferometry. 
For the present paper, we obtained mid-infrared interferometry of
S~Ori with VLTI/MIDI at four epochs in December 2004, February/March 2005, 
November 2005, and December 2005. We observed $v=1, J=1-0$ (43.1\,GHz) 
and $v=2, J=1-0$ (42.8\,GHz) SiO maser emission toward S~Ori with the VLBA 
in January, February, and November 2005, concurrent to within 5\% of the 
variability period to the first three VLTI/MIDI epochs. 
The first two common VLTI/VLBA epochs are located near visual minimum 
(phases 0.44 and 0.55) and the third after next visual maximum (phase 1.15). 
The fourth MIDI epoch occurred at a later post-maximum phase of 1.27.  

The mid-infrared MIDI data are sensitive to the structure
of the atmosphere consisting of the continuum photosphere and 
overlying molecular layers, as well as to 
the properties of the dust shell. The MIDI visibilities and flux densities
are modeled using the recent M model atmosphere series 
by Ireland et al. (\cite{ireland04b}), which is a dust-free self-excited 
dynamic model atmosphere series, to which we add an ad-hoc radiative 
transfer calculation of the dust shell. The constrained model parameters 
include the continuum photospheric angular diameter, as well as 
characteristics of the dust shell. 
The MIDI visibility and total flux data at all epochs can 
be described well by the chosen approach of combining the dust-free M model
atmosphere series with an ad-hoc radiative transfer model of the 
dust shell. Different model parameters can be constrained by different 
subsets of our MIDI data. 

The resulting continuum photospheric angular diameters at visual variability
phases 0.42, 0.55, 1.16, and 1.27 are 9.0\,$\pm$\,0.3\,mas,
7.9\,$\pm$\,0.1\,mas, 9.7\,$\pm$\,0.1\,mas, and 
9.5\,$\pm$\,0.4\,mas, respectively. The photospheric radius thus shows
a significantly phase-dependent size with amplitude of $\sim$\,20\% 
that is well correlated in phase with the visual lightcurve.

The dust shell can be modeled well with Al$_2$O$_3$ grains alone, and our
data show no indication of silicates, consistent with the modeling of the
IRAS LRS spectra by Lorenz-Martins \& Pompeia (\cite{lorenz00}). 
The inner dust shell angular diameters are 16.2\,$\pm$\,1.9\,mas,
15.9\,$\pm$\,1.6\,mas, 21.2\,$\pm$\,1.9\,mas, and 
22.8\,$\pm$\,2.1\,mas, respectively, for the visual phases given above,
corresponding to 1.8, 2.0, 2.2, and 2.4 times the continuum photospheric 
diameters. Thus, dust starts to form very close to the stellar surface,
corresponding to the region of the low-intensity wings of the 
atmospheric molecular layers of our dust-free atmosphere model.
In particular, the inner dust shell radius is smallest at the 
near-minimum phases, located directly outside of the steep decrease in the
dust-free, mid-infrared model intensity. The inner dust-shell boundary
expands to larger distances from the stellar surface at our post-maximum 
phases.
The overall Al$_2$O$_3$ optical depth, as well as the density gradient,
is largest at our near-minimum phases and significantly lower at the 
post-maximum stellar phases. These dust shell characteristics can be 
explained by a radiatively driven wind with increased mass-loss near 
visual stellar minimum occurring close to the stellar surface, 
and an expanded dust shell at the post-maximum phases.

The 43.1\,GHz and 42.8\,GHz maser spots show the typical structure of
partial to full rings with a clumpy distribution. There is no sign of a 
globally asymmetric gas distribution. 
The 43.1\,GHz mean maser ring angular diameters at visual phases 0.46, 0.56, 
and 1.14 are 19.6\,$\pm$\,2.8\,mas, 19.0\,$\pm$\,2.6\,mas, and 
20.4\,$\pm$\,2.4\,mas, respectively. The values for the 42.8\,GHz maser
are systematically smaller by 3--19\% and amount to 19.0\,$\pm$\,2.0\,mas, 
17.8\,$\pm$\,3.4\,mas, and 17.8\,$\pm$\,1.8\,mas. The indication of
systematically smaller mean maser ring diameters of the 42.8\,GHz transition 
compared to the 43.1\,GHz transition is consistent with earlier results 
in the literature on other Mira stars.
The velocity structure of both maser transitions indicate an expansion of 
the maser shell. This expansion is most likely an accelerated expansion, but
we cannot completely rule out a uniform expansion. The expansion velocity
for both transitions ranges between about 7\,km\,s$^{-1}$ and 
about 10\,km\,s$^{-1}$ between the inner and outer shell radii based 
on a scenario of an accelerated expansion or amounts 
to about 10\,km\,s$^{-1}$ 
based on a scenario with uniform expansion.

Compared to the continuum photospheric radius $R_\mathrm{Phot}$, 
the 43.1\,GHz/42.8\,GHz SiO maser ring radii are located at
2.2\,$\pm$0.3/2.1\,$\pm$0.2 (phase 0.44), 
2.4\,$\pm$0.3/2.3\,$\pm$0.4 (0.55), 
and 2.1\,$\pm$\,0.3/1.9\,$\pm$\,0.2 (1.15) times $R_\mathrm{Phot}$.
These values are consistent with theoretical estimates 
and earlier observations available in the literature.
Compared to the molecular atmosphere, the maser shells mark the region of
the molecular atmosphere just beyond the steepest decrease in the
mid-infrared model intensity.
Compared to the dust shell, the maser spots are co-located with the inner
dust shell near stellar minimum. At our post-maximum phases, the maser spots
remain at about the same location, while the dust shell has expanded outward.

Altogether, our observations indicate a pulsation of the stellar photosphere
approximately in phase with the visual lightcurve. Our measurements suggest
an increased mass-loss rate near stellar minimum and formation of 
Al$_2$O$_3$ dust grains occurring close to the stellar surface at about
1.8-2.0 photospheric radii directly outward of the dense molecular atmosphere, 
co-located with the extended low-intensity wings of the molecular 
atmosphere and with the SiO maser spots. At the post-maximum phases, 
the dust shell has expanded to about 2.2-2.4 photospheric radii, while 
the mean radius of the maser spots remains at $\sim$2.0 photospheric radii.  

Figure~\ref{fig:sketch} shows a sketch of S~Ori and its CSE near
visual minimum and post-maximum illustrating the findings of the present
work. 
Our results on S~Ori are largely consistent
with the canonical scheme of the CSE of AGB stars
discussed for instance in Reid \& Menten (\cite{reid97}). However, our 
results indicate an inner dust shell boundary that is located closer to the 
optical photosphere compared to the Reid \& Menten scheme
at about two photospheric radii, and SiO maser shells that
are co-located with the low-intensity molecular atmosphere {\it and} 
with Al$_2$O$_3$ dust grains near visual minimum. In the case of S~Ori,
the dust shell consists only of Al$_2$O$_3$ grains.
For Mira stars that also show a dust shell of silicates, the inner boundary 
of silicate grains is still expected at 
greater distances of $\ga$\,4 photospheric radii 
(Lorenz-Martins \& Pompeia \cite{lorenz00}). In addition to a static
picture of a Mira star's CSE, our observations suggest the formation
of a dust shell close to the surface near minimum phase and an expansion
by $\sim 20\%$ of the inner dust-shell boundary toward post-maximum phase.

Our modeling approach of adding the CLVs of a dust-free M model to
that of a radiative transfer model of the pure dust shell includes
two uncertainties. Firstly, the M model series is designed for 
the prototype Miras $o$~Cet and R~Leo and used here for the longer-period 
Mira S~Ori. Secondly, we omit a possible interaction of gas and dust 
outward of the inner dust shell boundary. Both effects might add some
additional uncertainty to the absolute values of the obtained photospheric
diameter values and dust shell characteristics. However, our data can
be described well by our modeling approach and these additional uncertainties 
are not expected to be significantly larger than the diameter errors 
given in Table~\ref{tab:results} and the step sizes of the dust shell 
parameters in Table~\ref{tab:chi2}. Furthermore, the relative accuracy
of the obtained parameters among the four epochs of observations are clearly
less affected by possible systematic uncertainties.
Currently, the chosen approach appears to be the best available one.
Similar studies of S~Ori and other long-period Mira stars in the future
would benefit from new dynamic model atmosphere series tailored to 
Mira stars with a longer period and a higher mass compared to $o$~Cet and
R~Leo, as well as from self-consistent calculations of dynamic 
atmospheres and dust formation.

Our mid-infrared interferometric data, in particular at the lower edge
of MIDI's wavelength range, together with the use of the M model series
provide constraints on the continuum photospheric angular diameter of S~Ori.
The resulting values are in good agreement with several earlier 
diameter measurements obtained in the near-infrared $K$-band by
different instruments and at different stellar phases, which increases
the confidence in our modeling approach. Additional 
confidence in the continuum photospheric diameters and the validity of the
dynamic model atmospheres can in future be reached by time series 
of high spectral resolution ($R=1000-10\,000$) near-infrared interferometry
with the VLTI/AMBER (Petrov et al. \cite{petrov03}) instrument 
concurrent to VLTI/MIDI observations.

Our results obtained at 4 epochs within one pulsation cycle suggest 
the formation of a dust layer consisting of Al$_2$O$_3$
grains close to the stellar surface near stellar minimum and an expanded
dust shell after stellar maximum. Our understanding of the connection
between the dust formation frequency and the stellar pulsation will be
further improved by using more epochs within one pulsation
cycle. Dust-forming episodes may also have their own timescale that 
may not be directly related to the visual variability 
(e.g., H\"ofner \& Dorfi \cite{hoefner97}; Nowotny et al. \cite{nowotny05}), 
a study of which requires observations extended over more than 
one pulsation cycle. Since the data of S~Ori do not show any 
sign of silicate grains, similar observations of targets that belong 
to the {\it intermediate class} group of Lorenz-Martins \& Pompeia 
(exhibiting Al$_2$O$_3$ {\it and} silicate grains) would be required to study 
the detailed relationship among the Al$_2$O$_3$ shell, the silicate shell, 
and the photospheric pulsation.  

Both our MIDI data and the images of the SiO maser shell
do not show  any signature of a globally asymmetric 
distribution of dust or gas in the case of S~Ori within 
our uncertainties. In general, asymmetric 
distributions of gas and dust
are to be expected for some Mira stars (e.g., Weigelt et al. \cite{weigelt96};
Boboltz \& Diamond \cite{boboltz05}; Tatebe et al. \cite{tatebe06}).
A detection of asymmetric structures can be performed at the VLTI by
sequentially using baselines of similar length but different orientation 
close in time/stellar phase to MIDI and by using the closure-phase 
instrument AMBER with baseline triangles comprised of baselines with 
different orientation. 

The relative positions of star and SiO maser shells in our current study
is unknown. Our comparison of the extension of the SiO maser shells to 
the continuum photospheric diameter is based on the assumption that the 
center of the maser spot distribution coincides with the center of the star.
This assumption is the most likely scenario for stars that do not show
any sign of asymmetric dust and gas distributions, such as S~Ori. In principle,
the relative position of the maser shells could be related relative to the
radio photosphere, which was detected by Reid \& Menten (\cite{reid97}) for
a sample of 6 Mira and semi-regular variables. Additional measurements 
of H$_2$O and OH masers relative to the SiO masers could give additional
information on the structure and kinematics of the more extended CSE. 
\begin{acknowledgements}
We thank Walter Jaffe, Rainer K\"ohler, Christian Hummel,
and others involved for making publicly available the MIDI data 
reduction software packages and tools {\tt EWS}, {\tt MIA}, and 
{\tt MyMidiGui}, 
as well as for helpful discussions on MIDI data reduction. 
We are grateful for support by the ESO DGDF.
The work of MS was supported by a grant of the Deutsche
Forschungsgemeinschaft on ``Time Dependence of Mira Atmospheres''.
We acknowledge with thanks the variable-star observations
from the AAVSO International Database contributed by observers
worldwide and used in this research. This research has made
use of the AFOEV and SIMBAD databases, operated at the CDS, France.
This research has made use of NASA's Astrophysics Data System.
\end{acknowledgements}

\begin{thebibliography}{}
%
\bibitem[1997]{begemann97} 
Begemann, B., Dorschner, J., Henning, T., et al.\ 1997, \apj, 
476, 199
%
\bibitem[1988]{beichmann88} 
Beichman, C.~A., Neugebauer, G., Habing, H.~J., 
Clegg, P.~E., \& Chester, T.~J.\ 1988, 
Infrared astronomical satellite (IRAS) catalogs and atlases.
%
\bibitem[1990]{benson90}
Benson, P.~J., Little-Marenin, I.~R., Woods, T.~C., 
et al.\ 1990, \apjs, 74, 911
%
\bibitem[1997]{boboltz97} 
Boboltz, D.~A., Diamond, P.~J., \& Kemball, A.~J.\ 1997, \apjl, 
487, L147 
%
\bibitem[2000]{boboltz00} 
Boboltz, D.~A., \& Marvel, K.~B.\ 2000, \apjl, 545, L149 
%
\bibitem[2005]{boboltz05b}
Boboltz, D.~A., \& Diamond, P.~J.\ 2005, \apj, 625, 978
%
\bibitem[2005]{boboltz05}
Boboltz, D.~A., \& Wittkowski, M.\ 2005, \apj, 618, 953 (BW05)
%
\bibitem[1994]{bujarrabal94}
Bujarrabal, V.\ 1994, \aap, 285, 953
%
\bibitem[1986]{chapman86}
Chapman, J.~M., \& Cohen, R.~J.\ 1986, \mnras, 220, 513
%
\bibitem[1996]{cho96}
Cho, S.-H., Kaifu, N., \& Ukita, N.\ 1996, \aj, 111, 1987
%
\bibitem[1999]{cohen99} 
Cohen, M., Walker, R.~G., Carter, B., et al.\ 1999, \aj, 
117, 1864 
%
\bibitem[2004]{cotton04}
Cotton, W.~D., Mennesson, B., Diamond, P.~J., et al.\ 2004, 
\aap, 414, 275
%
\bibitem[2006]{cotton06}
Cotton, W.~D., Vlemmings, W., Mennesson, B., et al.\ 2006, 
\aap, 456, 339
%
\bibitem[1994]{danchi94} 
Danchi, W.~C., Bester, M., Degiacomi, C.~G., Greenhill, L.~J., 
\& Townes, C.~H.\ 1994, \aj, 107, 1469
%
\bibitem[2000]{desmurs00}
Desmurs, J.~F., Bujarrabal, V., Colomer, F., \& Alcolea, J.\ 2006,
\aap, 360, 189
%
\bibitem[1994]{diamond94} 
Diamond, P.~J., Kemball, A.~J., Junor, W., et al.\ 1994, \apjl, 
430, L61
%
\bibitem[2003]{diamond03} 
Diamond, P.~J., \& Kemball, A.~J.\ 2003, \apj, 599, 1372
%
\bibitem[1995]{doel95}
Doel, R.~C., Gray, M.~D., Humphreys, E.~M.~L., Braithwaite, M.~F., \&
Field, D.\ 1995, \aap, 302, 797
%
\bibitem[1989]{feast89} 
Feast, M.~W., Glass, I.~S., Whitelock, P.~A., 
\& Catchpole, R.~M.\ 1989, \mnras, 241, 375
%
\bibitem[2005]{fedele05} 
Fedele, D., Wittkowski, M., Paresce, F., et al.\ 2005, \aap, 
431, 1019
%
\bibitem[2003]{glindemann03}
Glindemann, A., Algomedo, J., Amestica, R., et al.\ 2003, 
\procspie, 4838, 89
%
\bibitem[2000]{gray00} 
Gray, M.~D., \& Humphreys, E.~M.~L.\ 2000, New Astronomy, 5, 155 
%
\bibitem[1995]{haniff95} 
Haniff, C.~A., Scholz, M., \& Tuthill, P.~G.\ 1995, \mnras, 
276, 640
%
\bibitem[2006]{henden06}
Henden, A.~A.\ 2006, 
Observations from the AAVSO International Database, private 
communication.
%
\bibitem[1997]{hoefner97} 
H{\"o}fner, S., \& Dorfi, E.~A.\ 1997, \aap, 319, 648
%
\bibitem[1998]{hofmann98} 
Hofmann, K.-H., Scholz, M., \& Wood, P.~R.\ 1998, \aap, 339, 846 
%
\bibitem[2000]{hofmann00}
Hofmann, K.-H., Balega, Y., Scholz, M., \& Weigelt, G.\ 2000, \aap, 353, 1016
%
\bibitem[2001]{hofmann01} 
Hofmann, K.-H., Balega, 
Y., Scholz, M., \& Weigelt, G.\ 2001, \aap, 376, 518
%
\bibitem[2002]{hofmann02} 
Hofmann, K.-H., Beckmann, U., Bl\"ocker, T., et al.\ 2002, 
New Astronomy, 7, 9
%
\bibitem[2001]{hollis01} 
Hollis, J.~M., Boboltz, D.~A., Pedelty, J.~A., White, S.~M., \& 
Forster, J.~R.\ 2001, \apjl, 559, L37 
%
\bibitem[1996]{humphreys96} 
Humphreys, E.~M.~L., Gray, M.~D., Yates, J.~A., Field, D., 
Bowen, G., \& Diamond, P.~J.\ 1996, \mnras, 282, 1359 
%
\bibitem[2002]{humphreys02} 
Humphreys, E.~M.~L., Gray, M.~D., Yates, J.~A., Field, D., 
Bowen, G.~H., \& Diamond, P.~J.\ 2002, \aap, 386, 256
%
\bibitem[2006]{ireland06} 
Ireland, M.~J., \& Scholz, M.\ 2006, \mnras, 367, 1585
%
\bibitem[2004a]{ireland04a} 
Ireland, M.~J., Scholz, M., \& Wood, P.~R.\ 2004a, \mnras, 
352, 318
%
\bibitem[2004b]{ireland04b}
Ireland, M.~J., Scholz, M., Tuthill, P.~G., \& 
Wood, P.~R.\ 2004b, 
\mnras, 355, 444
%
\bibitem[2005]{ireland05}
Ireland, M.~J., Tuthill, P.~G., Davis, J., \& Tango, W.\ 2005, 
\mnras, 361, 337
%
\bibitem[1997]{dusty1} 
Ivezi{\'c}, {\v Z}., \& Elitzur, M.\ 1997, \mnras, 287, 799
%
\bibitem[1999]{dusty2}
Ivezi{\'c}, {\v Z}., Nenkova, M., \& Elitzur, M\ 1999, User Manual for {\tt DUSTY},
University of Kentucky Internal Report, accessible 
at http://www.pa.uky.edu/\~moshe/dusty
%
\bibitem[2002]{jacob02}
Jacob, A.~P., \& Scholz, M.\ 2002, \mnras, 336, 1377
%
\bibitem[2003]{jeong03} 
Jeong, K.~S., Winters, J.~M., Le Bertre, T., \& Sedlmayr, E.\ 
2003, \aap, 407, 191
%
\bibitem[1991]{jewell91}
Jewell, P.~R., Snyder, L.~E., Walmsley, C.~M., Wilson, T.~L., \&
Gensheimer, P.~D.\ 1991, \aap, 242, 211
%
\bibitem[1990]{jura90} 
Jura, M., \& Kleinmann, S.~G.\ 1990, \apjs, 73, 769
%
\bibitem[1997]{kemball97} 
Kemball, A.~J., \& Diamond, P.~J.\ 1997, \apjl, 481, L111
%
\bibitem[1995]{koike95}
Koike, C., Kaito, C., Yamamoto, T., et al.\ 1995, \icarus, 
114, 203
%
\bibitem[2003]{leinert03} 
Leinert, Ch., Graser, U., Richichi, A., et al.\ 2003, 
The Messenger, 112, 13
%
\bibitem[2004]{leinert04} 
Leinert, Ch., van Boekel, R., Water, L.~B.~F.~M., et al.\ 2004, 
\aap, 423, 537
%
\bibitem[1992]{lockett92}
Lockett, P., \& Elitzur, M.\ 1992, \apj, 399, 704
%
\bibitem[2000]{lorenz00}
Lorenz-Martins, S., \& Pompeia, L.\ 2000, \mnras, 315, 856
%
\bibitem[2002]{mennesson02} 
Mennesson, B., Perrin , G., Chagnon, G., et al.\ 2002, \apj, 
579, 446
%
\bibitem[2005]{millan05}
Millan-Gabet, R., Pedretti, E., Monnier, J.~D., et al.\ 2005, 
\apj, 620, 961
%
\bibitem[2005]{nowotny05} 
Nowotny, W., Aringer, B., H{\"o}fner, S., Gautschy-Loidl, R., 
\& Windsteig, W.\ 2005, \aap, 437, 273 
%
\bibitem[2005]{ohnaka05}
Ohnaka, K., Bergeat, J., Driebe, T., et al.\ 2005, \aap, 429, 1057 
%
\bibitem[2006a]{ohnaka06a}
Ohnaka, K., Driebe, T., Hofmann, K.-H., et al.\ 2006a, \aap, 45, 1015
%
\bibitem[2006b]{ohnaka06b}
Ohnaka, K., Scholz, M., \& Wood, P.~R\ 2006b, \aap, 446, 1119
%
\bibitem[2007]{ohnaka07}
Ohnaka, K., Driebe, T., Weigelt, G., \& Wittkowski, M.\ 2007, \aap, 466, 1099
%
\bibitem[1992]{ossenkopf92}
Ossenkopf, V., Henning, Th., \& Mathis, J. S.\ 1992, A\&A, 261, 567
%
\bibitem[2004]{perrin04} 
Perrin, G., Ridgway, S.~T., Mennesson, B., et al.\ 2004, \aap, 
426, 279
%
\bibitem[2003]{petrov03}
Petrov, R., Malbet, F., Weigelt, G., et al.\ 2003, \procspie, 4838, 924
%
\bibitem[1992]{quirrenbach92} 
Quirrenbach, A., Mozurkewich, D., Armstrong, J.~T., 
et al.\ 1992, \aap, 259, L19
%
\bibitem[1977]{reid77}
Reid, M.~J., Muhleman, D.~O., Moran, J.~M., Johnston, K.~J., \&
Schwartz, P.~R.\ 1977, \apj, 214, 60
%
\bibitem[1997]{reid97}
Reid, M.~J., \& Menten, K.~M.\ 1997, \apj, 476, 327 
%
\bibitem[2004]{samus04} 
Samus, N.~N., Durlevich, O.~V., \& et al.\ 2004,
Combined General Catalog of Variable Stars (GCVS4.2, 2004 Ed.) 
VizieR Online Data Catalog, 2250, 0
%
\bibitem[2001]{scholz01}
Scholz, M.\ 2001, \mnras, 321, 347
%
\bibitem[2003]{scholz03}
Scholz, M.\ 2003, \procspie, 4838, 163
%
\bibitem[1998]{sloan98} 
Sloan, G.~C., \& Price, S.~D.\ 1998, \apjs, 119, 141
%
\bibitem[2006]{tatebe06}
Tatebe, K., Chandler, A.~A., Hale, D.~D.~S., \& Townes, C.~H.\ 2006, 
\apj, 652, 666
%
\bibitem[2003a]{tej03a}
Tej. A., Lan{\c c}on, A., Scholz, M.\ 2003a, \aap, 401, 347
%
\bibitem[2003b]{tej03b} 
Tej, A., Lan{\c c}on, A., Scholz, M., \& Wood, P.~R.\ 2003b, 
\aap, 412, 481
%
\bibitem[2005]{templeton05}
Templeton, M.~R., Mattei, J.~A., \& Willson, L.~A.\ 2005, \aj, 
130, 788
%
\bibitem[2002]{thompson02} 
Thompson, R.~R., Creech-Eakman, M.~J., \& van Belle, 
G.~T.\ 2002, \apj, 577, 447
%
\bibitem[1996]{vanbelle96}
van Belle, G.~T., Dyck, H.~M., Benson, J.~A., \& Lacasse, 
M.~G.\ 1996, \aj, 112, 2147
%
\bibitem[2002]{vanbelle02}
van Belle, G.~T., Thompson, R.~R., \& 
Creech-Eakman, M.~J\ 2002, \aj, 124, 1706
%
\bibitem[1996]{weigelt96} 
Weigelt, G., Balega, Y., Hofmann, K.-H., \& Scholz, M.\ 1996, \aap, 316, L21
%
\bibitem[2006]{weiner06} 
Weiner, J., Tatebe, K., Hale, D.~D.~S. et al.\ 2006, \apj, 
636, 1067
%
\bibitem[1953]{wilson53}
Wilson, R.~E.\ 1953, ``General catalog of stellar radial velocities'',
Carnegie Institute Washington D.C.
%
\bibitem[2003]{winters03}
Winters, J.~M., Le Bertre, T., Jeong, K.~S., Nyman, L.-A., 
\& Epchtein, N.\ 2003, \aap, 409, 715
%
\bibitem[2006]{wittkowski06} 
Wittkowski, M., Aufdenberg, J.~P., Driebe, T., 
et al.\ 2006, \aap, 460, 855
%
\bibitem[2006]{woitke06}
Woitke, P.\ 2006, \aap, 460, L9
%
\bibitem[2004]{woodruff04} 
Woodruff, H.~C., Eberhardt, M., Driebe, T., 
et al.\ 2004, \aap, 421, 703
%
\bibitem[1983]{wyatt83}
Wyatt, S.~P., \& Cahn, J.~H.\ 1983, \apj, 275, 225
%
\bibitem[1994]{yates94}
Yates, J.~A., \& Cohen, R.~J.\ 1994, \mnras, 270, 958
%
\bibitem[2005]{yi05} 
Yi, J., Booth, R.~S., Conway, J.~E., \& Diamond, P.~J.\ 2005, 
\aap, 432, 531 
%
\bibitem[1995]{young95}
Young, K.\ 1995, \apj, 445, 872
%
\end{thebibliography}
\end{document}